\documentclass[preprint, authoryear,10pt]{elsarticle}

\usepackage{lineno,hyperref}
\modulolinenumbers[5]

\usepackage{amsmath,amsfonts,amssymb,graphicx,amsthm}
\usepackage{graphicx}
\usepackage{graphicx}
\usepackage{epstopdf}
\usepackage{hyperref} 
\usepackage{color}
\usepackage[normalem]{ulem}

\usepackage{float}
\usepackage{caption}
\usepackage{subcaption}
\usepackage{epstopdf} 
\usepackage{booktabs}
\usepackage{longtable,array,booktabs,multirow}
\usepackage{bbm}
\usepackage{eso-pic}

\theoremstyle{definition}

\newcommand{\be}{\begin{equation}}
\newcommand{\ee}{\end{equation}}
\newcommand{\bea}{\begin{eqnarray}}
\newcommand{\eea}{\end{eqnarray}}
\newcommand{\beas}{\begin{eqnarray*}}
\newcommand{\eeas}{\end{eqnarray*}}

\newtheorem{Def}{Definition}

\biboptions{authoryear}

\begin{document}

\begin{frontmatter}

\title{Taming the Basel Leverage Cycle}

\author[INET,OX]{Christoph Aymanns}\ead{aymanns@maths.ox.ac.uk}
\author[UCL,SRC]{Fabio Caccioli}\ead{f.caccioli@ucl.ac.uk}
\author[INET,OX,SFI]{J. Doyne Farmer}\ead{Doyne.Farmer@inet.ox.ac.uk}
\author[OX]{Vincent W.C. Tan}\ead{vintwc@gmail.com}

\address[INET]{Institute of New Economic Thinking at the Oxford Martin School, University of Oxford, Oxford OX2 6ED, UK}
\address[OX]{Mathematical Institute, University of Oxford, Oxford OX1 3LB, UK}
\address[UCL]{Department of Computer Science, University College London, London, WC1E 6BT, UK}
\address[SRC]{Systemic Risk Centre, London School of Economics and Political Sciences, London, UK}
\address[SFI]{Santa Fe Institute, Santa Fe, NM 87501, USA}

\begin{abstract}

Effective risk control must make a tradeoff between the microprudential risk of exogenous shocks to individual institutions and the macroprudential risks caused by their systemic interactions.  We investigate a simple dynamical model for understanding this tradeoff, consisting of a bank with a leverage target and an unleveraged fundamental investor subject to exogenous noise with clustered volatility.   The parameter space has three regions: (i) a stable region, where the system always reaches a fixed point equilibrium; (ii) a locally unstable region, characterized by cycles and chaotic behavior; and (iii) a globally unstable region.  A crude calibration of parameters to data puts the model in region (ii).  In this region there is a slowly building price bubble, resembling a ``Great Moderation", followed by a crash, with a  period of approximately 10-15 years, which we dub the {\it Basel leverage cycle}.  We propose a criterion for rating macroprudential policies based on their ability to minimize risk for a given average leverage.  We construct a one parameter family of leverage policies that allows us to vary from the procyclical policies of Basel II or III, in which leverage decreases when volatility increases, to countercyclical policies in which leverage increases when volatility increases.  We find the best policy depends critically on three parameters: The average leverage used by the bank; the relative size of the bank and the fundamentalist, and the amplitude of the exogenous noise. Basel II is optimal when the exogenous noise is high, the bank is small and leverage is low; in the opposite limit where the bank is large or leverage is high the optimal policy is closer to constant leverage.  We also find that systemic risk can be dramatically decreased by lowering the leverage target adjustment speed of the banks. 
\end{abstract}


\end{frontmatter}

\tableofcontents

\section{Introduction} 
Borrowing in finance is often called ``leverage", which is inspired by the fact that borrowing increases returns, much as a mechanical lever makes it possible to increase forces.   But leverage increases not only return but also risk, which naturally motivates lenders to introduce constraints on its use.\footnote{Such a constraint may arise in a number of ways. If the investor is using collateralized loans to fund its investments, it must maintain margin on its collateral. Alternatively, a regulator may impose a risk contingent capital adequacy ratio. Finally, internal risk management considerations may lead the investor to adopt a Value-at-Risk constraint.  (In simple terms Value-at-Risk is a measure of how much the bank could lose with a given small probability).  All of these cases effectively impose a risk contingent leverage constraint.}    

Because leverage goes up when prices go down, a drop in prices tightens leverage constraints, which often forces investors to sell into falling markets.\footnote{In principle, distressed banks can reduce their leverage in two ways: they can raise more capital or sell assets. In practice most banks tend to do the latter, as documented in \cite{Adrian2008a}.} 
As investors sell into falling markets they cause prices to fall further.  This triggers a positive feedback loop in which selling depresses prices, which causes further selling, which further tightens leverage constraints, etc.  Similarly, positive news about prices causes a decline in perceived risk, which loosens leverage constraints, which causes further price increases.   These dynamics were termed the {\it leverage cycle} by Geanakopolos.\footnote{Minsky was the first author that we know of to describe the leverage cycle in qualitative terms.  The first quantitative model is due to \cite{Geanakoplos1997,Geanakoplos2003,Fostel2008,Geanakoplos2010}.  See also the early model by \cite{Gennotte1990}, which models the destabilizing effects of leverage but does not model the cycle per se.  Another relevant model is \cite{Brunnermeier2008a}, where the authors investigate the destabilizing feedback between funding liquidity and market liquidity. A further discussion on the destabilizing effects of margin can be found in \cite{Gorton2010}.} Constraining the use of leverage is clearly beneficial at the individual level. At the systemic level, however, the dynamics induced by leverage constraints can lead to booms and busts; it is widely believed that excessive leverage caused or at least exacerbated the recent financial crisis.

\subsection{Empirical motivation}\label{SEC::Empirical}
\begin{figure}
\centering
\includegraphics[width=0.95\textwidth]{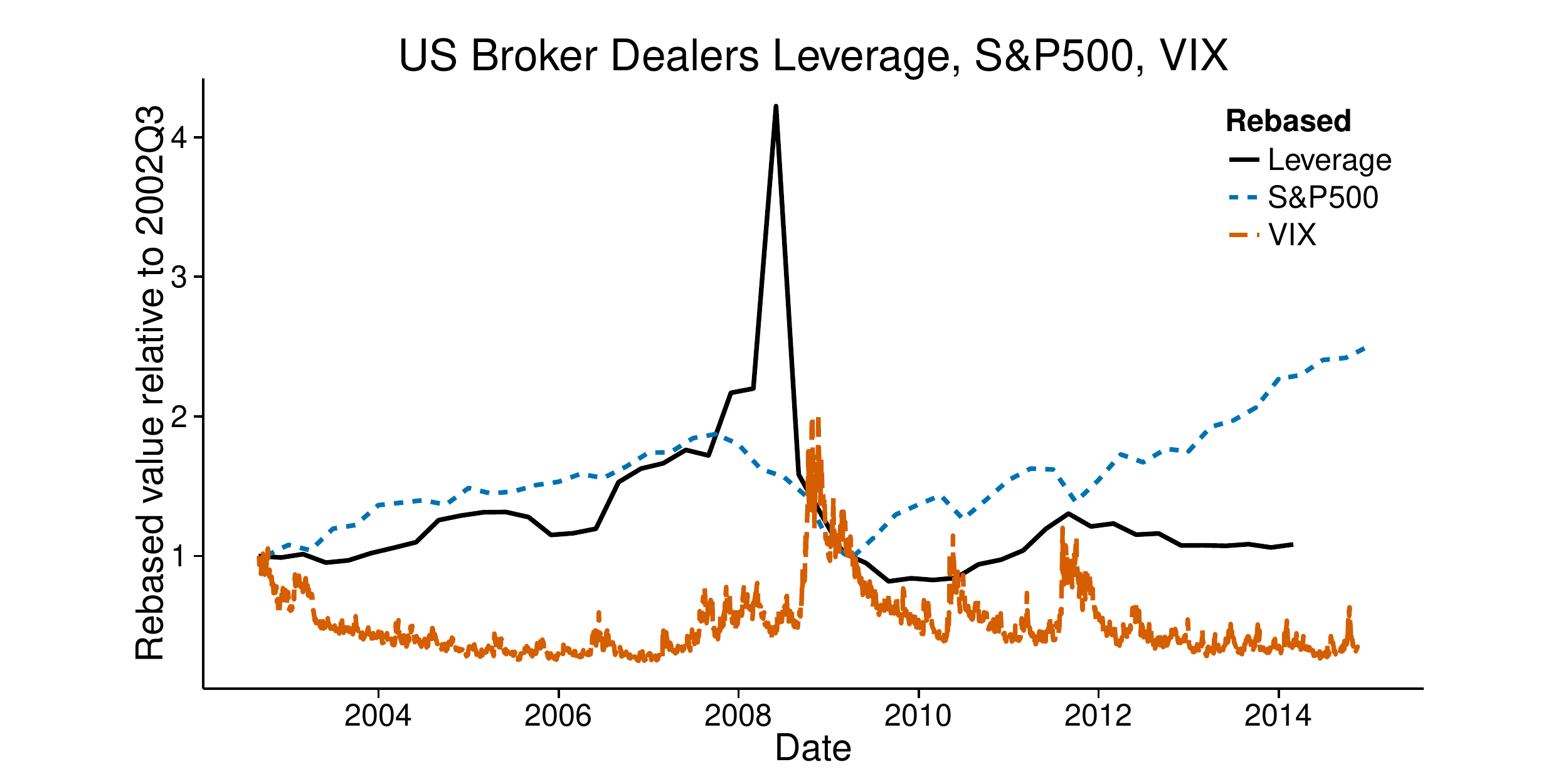}
 \caption{The leverage of US Broker-Dealers (solid black line) compared to the S\&P500 index (dashed blue line) and the VIX S\&P500 (red dash-dotted line.  Data is quarterly; see footnote 7.} 
         \label{FIG::data_leverage_cycle}
\end{figure}
The period encompassing the Great Moderation and the subsequent global financial crisis starting in late 2007 is a case in point for the strong correlation between leverage, market volatility and asset prices. Detailed evidence over a longer time horizon for the link between asset prices and leverage for various types of financial institutions is provided by \cite{Adrian2010}. In this section we focus only on the Great Moderation and the subsequent financial crisis.

During the Great Moderation perceived volatility, as measured by the VIX index of expected future volatility, declined consistently over several years, as shown by the dotted line in Figure \ref{FIG::data_leverage_cycle}.   At the same time, in a near mirror image to volatility, asset prices (as measured by the S\&P500 index) and leverage of financial institutions (as measured by the leverage of US security broker dealers) consistently increased.\footnote{
It should be noted that US security broker dealers are a somewhat extreme example of leveraged financial institutions and are not representative for the behavior of commercial banks. Here we use their example to illustrate the stark correlation between leverage, volatility and asset prices in an anecdotal way. A more nuanced evaluation can be found in \cite{Adrian2010}.  The data on US Security Broker Dealer Leverage, (defined as Assets/(Assets-Liabilities), is from US Federal Reserve Flow of Funds Data Package F.128 available at \texttt{http://www.federalreserve.gov/datadownload/}. }  
As financial institutions expanded their leverage their assets and liabilities grew correspondingly. 

The Great Moderation came to a sudden and dramatic halt when the US subprime mortgage crisis began to unfold in late 2007 and the subsequent financial crisis sent asset prices into a downward spiral.  As asset prices began to fall, market volatility increased. At the same time the leverage of financial institutions increased -- in fact drastically so. This was due the fact that even relatively small drops in asset prices can massively increase the leverage of financial institutions that are already highly leveraged. Financial institutions responded quickly to increased market volatility and deleveraged by a factor of 2 in the span of one quarter.

This deleveraging likely had a drastic negative impact on asset prices.  Of course, many other factors affected asset prices at that time, and this evidence is only anecdotal.  Nonetheless, the correlation observed between asset prices, volatility and leverage motivates the model that is developed here, which is designed to address the following questions:  How do leveraged investors respond to changes in market prices? How are market prices affected by investors' portfolio adjustments? How does this potential feedback loop affect the overall dynamics of the financial system? Finally, what should regulators do to control this feedback loop, and how should they make an appropriate compromise between microprudential and macroprudential regulation? 

Our model suggests that the underlying cause of these events might have been due to the simple combination of leverage and dynamic risk management in the style of Basel II, and that the collapse of the housing bubble might have been only the spark that happened to cause the crash.  Of course the real situation was complicated and this cannot be proven, but our model makes this explanation plausible.

\subsection{Review of literature on leverage cycles}

There are many possible mechanisms that have been conjectured to drive leverage cycles.  In the original model of Geanakoplos investor heterogeneity plays a key role:  The most optimistic investors are also the most leveraged investors.   The leveraged investors are hit harder by downturns, which reduces their market power, which negatively impacts average expectations and amplifies downward price movements.  Many other factors have been conjectured to drive leverage cycles, including short-termism, herding and incentive distortions.\footnote{See for example \cite{Aikman2012}, \cite{deNicolo2012} and \cite{Gennaioli2012}.}  

{Here we focus on the side-effects of risk management as a driver of leverage cycles. A passive investor, i.e. an investor that never rebalances his or her portfolio, is {\it countercyclical} in the sense that falling prices drive leverage up and vice versa.  In contrast, \cite{Adrian2008a} point out that many investors, such as commercial banks, use constant leverage targets, creating a positive feedback between the demand for an asset and its return.  Since falling prices increase leverage, maintaining a constant target leverage causes investors to sell into a falling market and to buy into a rising market. Such behavior is inherently destabilizing: Higher (lower) demand leads to higher (lower) prices, that further increase (reduce) demand, and so on.

\cite{Adrian2008a} document even more destabilizing behavior by investors such as investment banks.   These investors are actively {\it procyclical}, i.e. they lower leverage targets when prices fall and raise them when prices rise.  This further amplifies the potentially destabilizing positive feedback between demand and returns.  Adrian and Shin argue that this can be due to regulatory risk management, since a risk neutral investor subject to a Value-at-Risk constraint increases her leverage when volatility is low and reduces it when volatility is high. 
In fact, volatility and prices cannot be disconnected.  There is empirical evidence for a negative correlation between returns and volatility \citep{Black76,Christie82,Nelson91,Engle93}. This implies that when prices increase (decrease), volatility decreases (increases) and target leverage goes up (down), which results in leverage procyclicality. The fact that minimum capital requirements based on VaR are likely to result in procyclical behavior was also pointed out by \cite{Estrella2004}.

Since VaR risk management induces the procyclicality of leverage with respect to prices, in the following we will refer to a {\it procyclical leverage control policy} as one for which banks are required  to reduce their target leverage when volatility increases, and are allowed to increase it when volatility decreases. It is important to stress that this feedback can have significant macro-economic consequences. For instance \cite{Heuvel2002} investigates the effect of capital regulation on the transmission of monetary policy via bank lending, finding that leverage procylicality can lead to an amplification of monetary policy. 

In contrast to the procyclical case, we will refer to a {\it countercyclical leverage control policy} as one for which banks can target a higher leverage if volatility is high, but are required to reduce their leverage when volatility is low. The rationale for such a policy is to counteract the potentially destabilizing positive feedback between demand and returns that occurs under procyclical policies.  We show that an actively countercyclical policy can also be destabilizing.  In fact, under a policy that is countercyclical with respect to risk, we observe in our model that the relation between returns and volatility can be reversed, and that higher (lower) volatility is associated with higher (lower) prices. Therefore, if an investor buys when volatility increases, this can further increase volatility and prices, which raises leverage targets, etc., also resulting in a positive feedback loop.  The challenge for policy makers is to find a leverage control policy that avoids the Scylla and Charybdis of excessively procyclical policies on one side or excessively countercyclical policies on the other.  The aim of this paper is to understand how the cyclicality of leverage control policies affects the properties of the financial system, and to seek the proper compromise between these two extremes.}

It has to be stressed that the concept of cyclicality we refer to in this paper is with respect to risk, not with respect to the behavior of macroeconomic indicators. For example, \cite{Drehmann2012} provide  counterfactual simulations showing how  leverage control policies that are countercyclical with respect to the difference between the credit-to-GDP ratio and its long-run average can help making the economy more stable. The focus of our paper, however, is on the circumstances in which risk control can cause financial instability, and how to make an effective tradeoff between systemic vs. individual risk.

The consequences of procyclical leverage have now been studied by many authors.\footnote{See for example \cite{Danielsson2001}, \cite{Danielsson2004}, \cite{Adrian2008a}, \cite{Shin2010}, \cite{Zigrand2010}, \cite{Colla2011}, \cite{Tasca2012}, \cite{Poledna2013}, \cite{Adrian2014} and \cite{Brummitt2014}.} The fact that feedback loops due to capital requirement constraints can lead to the amplification of shocks has been demonstrated, for example, by \cite{Zigrand2010}, \cite{He01042012},
 \cite{Thurner2010} and \cite{Adrian12b,Adrian13}. \cite{Aymanns2014} go beyond this by showing how leverage constraints can lead to an endogenous cycle, i.e. one in which spontaneous oscillations occur.  We call this the {\it Basel leverage cycle}.\footnote{However bear in mind that this can occur even without any regulatory policy constraint, due to prudent risk managers who limit the risk of individual institutions in isolation while failing to properly take systemic risk into account.}
While \cite{Adrian12b,Adrian13} study this through a dynamic stochastic general equilibrium model, \cite{Aymanns2014} consider a more stylized setting where investment decisions of households and unleveraged funds are not explicitly modeled.  The latter model has the virtue of being very simple, and also of showing how under bounded rationality endogenous leverage cycles can occur even in a deterministic limit where there are no shocks.   In this paper we modify the model of \cite{Aymanns2014} by adding a fundamental noise trader subject to exogenous noise, which allows us to study the tradeoff between micro and macro prudential regulation.  We also present a full stability analysis.

\subsection{Summary of our contribution}

Our main contribution is the identification of optimal leverage control policies.
As in the Aymanns and Farmer model, the exponent of the relationship between perceived risk and target leverage is a free parameter.  In this way it is possible to capture both procyclical and countercyclical leverage control policies in a single model. The ability to interpolate between procyclical and countercyclical leverage control policies in one model allows us to compare policies of different ``cyclicality'' and study their effectiveness in controlling leverage cycles.

In order to do this it is necessary to choose a criterion for selecting the optimal policy.  We believe a good policy is one that maximizes leverage at a given level of overall risk to the financial system.  Maximizing leverage is desirable because this means that the capital of the financial system is put to full use in providing credit to the real economy.   In fact, for reasons of convenience it is more feasible for us to minimize risk at a given leverage; this is essentially equivalent, since the average leverage can be adjusted to match any desired risk target (and the policy that is selected will be the same).  We measure risk in terms of realized shortfall, i.e. the average of large losses to the financial system as a whole. 

The main result of this paper is that the optimal policy 
depends critically on three parameters:  the average leverage used by the bank;  the relative size of the bank and the fundamentalist; and the amplitude of the exogenous noise. A procyclical leverage control policy such as that of Basel II is optimal when the exogenous noise is high, the volatility is strongly clustered, the bank is small and leverage is low; in the opposite limit where these conditions are not met the optimal policy is closer to constant leverage.

\section{A simple model of leverage cycles}

\subsection{Sketch of the model}

We consider a financial system composed of a leveraged investor (called a {\it bank} for simplicity), an unleveraged fund investor, which we call the {\it fund}, and a passive outside lender that provides credit as required by the bank.  The bank and the fund make a choice between investing in a risky asset whose price is determined endogenously vs. a risk free asset with fixed price, which we will call {\it cash}.

We focus on risk management by assuming the bank holds the relative weight of the risky asset and cash fixed. The bank's risk management consists of two components:  First, we assume that the bank estimates the future volatility of its investment in the risky asset by using an exponential moving average of historical returns.   Second, the bank uses the estimated volatility to set its desired leverage.  The target can be set either by internal risk management or by externally imposed regulatory constraints:  The net result is the same.  If the bank is below its desired leverage, it will borrow more and use the additional funds to expand its balance sheet.  Conversely, if it is above its desired leverage, it will liquidate part of its investments and pay back part of its debt.

The fund is a proxy for the rest of the financial system, i.e. the part that does not do leverage targeting.  Leverage targeting creates inherently unstable dynamics, as it implies buying into rising markets and selling into falling markets.  Thus it is necessary have at least one other investor who plays a stabilizing role.  We model the fund as a weakly fundamentalist investor whose investment decisions are perturbed by exogenous random shocks reflecting information flow or decision processes outside the scope of the model.  The exogenous random shocks display clustered volatility.  The fund investor and the bank interact through the market for the risky asset and the market clearing price is determined by the investments of the fund and the bank.    

Because of the clustered volatility, in the absence of any systemic effects, the bank should adjust its leverage to maintain a constant Value at Risk.  With systemic effects this becomes more complicated, which is what this model allows us to investigate.

In addition to its portfolio management decisions, we assume that the bank tries to maintain a constant target equity. This is consistent with the empirical observation that the equity of commercial and investment banks is roughly constant over time; see \cite{Adrian2008a}. In order to conserve cash flow in our model, we assume that dividends paid out by the bank when the equity exceeds the target are invested in the fund, while new capital invested in the bank when the equity is below the target is withdrawn from the fund. This prevents all the wealth from accumulating with either the bank or the fund and makes the asymptotic dynamics stationary. 

\begin{figure}
\centering
\includegraphics[width=0.95\textwidth]{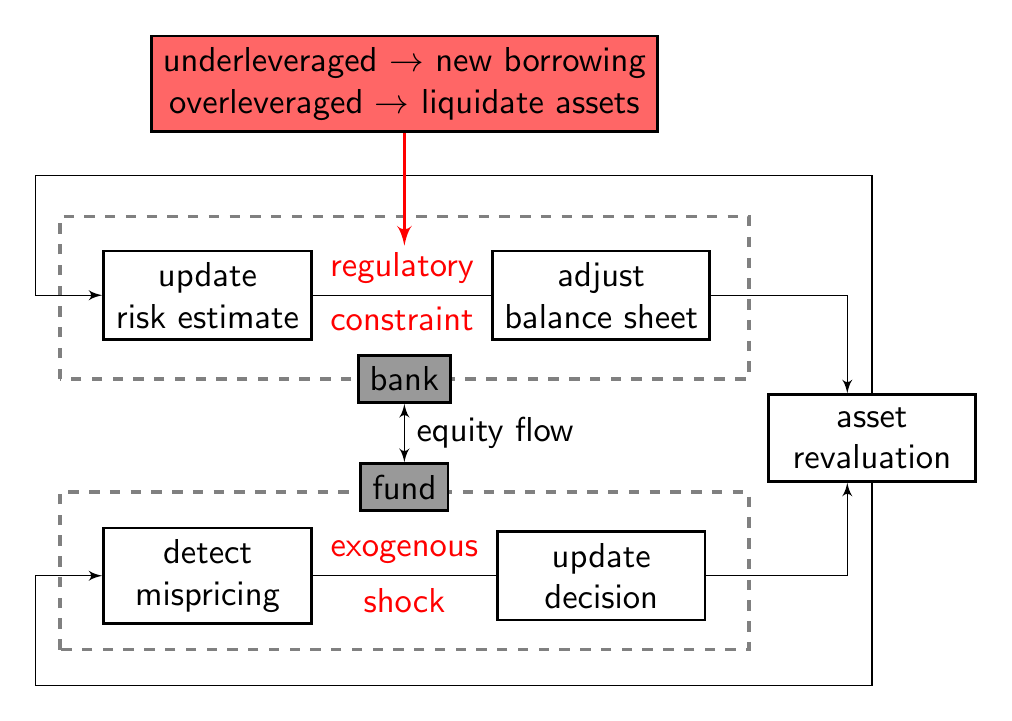}
 \caption{Diagrammatic representation of the model: The bank and the fund interact through price formation. The bank's demand for the risky asset depends on its estimated risk based on historical volatility and on its capital requirement. The demand of the fund consists of a mean reverting component that tends to push the price towards its fundamental value; in addition there is a random exogenous shock to the fund's demand that has clustered volatility.  Price adjustments affect the bank's estimation of risk and the mean reverting behavior of the fund. The cash flow consistency in the model is enforced by equity flowing between the bank and the fund in equal amounts. The driver of the endogenous dynamics is the feedback loop between price changes, volatility and demand for the risky asset.} 
         \label{FIG::diagram}
\end{figure}

Figure \ref{FIG::diagram} shows a diagrammatic representation of the model. The main driver of the dynamics is the feedback loop between changes in the price of the risky asset and balance sheet adjustments:  The bank reacts to price changes to maintain its capital requirements under its perception of risk; similarly the fund invests as one expect from a fundamentalist, buying when prices are below value and selling when they are above.  The balance sheet adjustments of the bank and the fund determine the price, which in turn feeds back to determine their decisions. 

\subsection{Leverage regulation} The most important ingredient of our model is the fact that the bank is subject to a capital requirement policy.
The leverage ratio\footnote{We use this definition of leverage in analogy to the Tier 1 regulatory leverage ratio (Tier 1 capital over bank total assets). An alternative definition of the leverage ratio only considers risky assets in the numerator. Since in our model the bank holds the share of risky assets to total assets fixed, this alternative definition simply introduces a multiplicative constant into the leverage calculation and does not affect the qualitative outcome of our model.} of the bank is defined as
\be\label{EQ::leverage}
\lambda(t) = \frac{\text{Total Assets}}{\text{Equity}},
\ee
and the capital requirement policy implies a leverage constraint of the type $\lambda(t)\le \bar{\lambda}(t)$, i.e. the bank is allowed a maximum leverage $\bar\lambda(t)$. A cap on leverage is equivalent to the existence of a minimum capital buffer that the bank is required to keep to absorb losses.
Conditional on the leverage constraint, it can be shown that the return on equity of the bank is maximized if $\lambda(t)=\bar\lambda(t)$ (see for instance \cite{Shin2010}).  We therefore assume that the bank always targets its maximum allowed leverage $\bar\lambda(t)$.\footnote{
In reality banks usually keep more capital than required by regulation in order to reduce the cost of recapitalization or portfolio adjustments associated with violation of the minimal capital requirement. Using this perspective, \cite{Peura2006} explain the pattern of capital buffers observed in a sample of US commercial banks. However, note that our results remain valid even if we assume that banks hold more capital than required by the regulator. We only require that the resulting bank capital buffer responds to changes in perceived risk in a well defined way, i.e. in our model changes in the capital buffer are more important than the level of the capital buffer.} 

We assume that  $\bar\lambda(t)$ depends on the bank's estimate of the volatility of the risky asset, i.e. $\bar\lambda(t) = F(\sigma^2(t))$, where $\sigma^2(t)$ is the bank's perceived risk.   Although nothing we do here depends on this, to gain intuition it is useful to compute the function $F$ under the special case of a Value-at-Risk constraint with normally distributed returns.  In this case the bank's target leverage is given by  (see for example \cite{Corsi2013}):
\begin{equation*}
\bar{\lambda}(t) = F_{\text{VaR}}(\sigma^2(t)) = \frac{1}{\sigma(t) \Phi^{-1} (a)} \propto \frac{1}{\sigma(t)},
\end{equation*}
where $\Phi$ is the cumulative distribution of the standard normal, $a$ is the VaR quantile, and $\sigma$ the volatility of the risky asset.
Under this specification the bank increases its leverage when the volatility  of the risky asset diminishes and decreases its leverage in the opposite case. 
Motivated by \cite{Adrian2014}, we classify leverage policies as follows:
\begin{Def}\label{Def:cyclicality}
A leverage policy $F(\sigma^2(t))$ is {\it procyclical} if $dF/d\sigma^2 < 0$ and {\it countercyclical} if $dF/d\sigma^2 > 0$.\footnote{
This definition could be generalized for any risk measure; we use the standard deviation $\sigma$ for simplicity.}
\end{Def}
A class of leverage control policies that allows us to interpolate between procyclical and countercyclial leverage control policies is given by
\begin{equation}\label{EQ::leverage_policy}
\bar{\lambda}(t) = F_{(\alpha,\sigma^2_0,b)}(\sigma(t)) := \alpha ( \sigma^2(t) + \sigma_0^2 )^b,
\end{equation}
where $\alpha > 0$, $\sigma^2_0 > 0$ and $b \in [-0.5,0.5]$. We refer to $\alpha$ as the bank's riskiness. The larger $\alpha$ the larger the bank's target leverage for a given level of perceived risk $\sigma^2(t)$.\footnote{
Note that under standard Value-at-Risk the bank's leverage depends on the variance of its entire portfolio which in our model includes non risky cash holdings. Usually, the portfolio variance is computed as the inner product of the covariance matrix with the portfolio weights. In our case this implies that the portfolio variance is simply $\sigma^2(t)$ scaled by the bank's investment weight in the risky asset $w_\text{B}$. However, since we take $w_B$ constant throughout, the resulting risk rescaling factor can be absorbed into $\alpha$ without loss of generality. Therefore, we make $F_{(\alpha,\sigma^2_0,b)}$ only a function of $\sigma(t)$.
}
For the special case where returns are normal with $b = -0.5$ and $\sigma_0^2 = 0$,  $\alpha$ is linked to the quantile used to measure VaR by the inverse cumulative normal distribution.  In the more general case where there are heavy tails or with other choices of parameters this correspondence is no longer valid.  However, since the variance of returns is finite, under the Chebyshev's bound VaR is bounded by a quantity inversely proportional to the standard deviation of the return distribution, so the conjectured relationship in Equation~(\ref{EQ::leverage_policy}) remains qualitatively correct.  The relationship between $\alpha$ and risk remains monotonic under any sensible risk measure, and one can simply think of $\alpha$ as a risk parameter and Equation\ (\ref{EQ::leverage_policy}) as a particular choice of $F$, corresponding to a volatility estimate based on historical standard deviation.

\begin{figure}[h]
\centering
\includegraphics[width=0.70\textwidth]{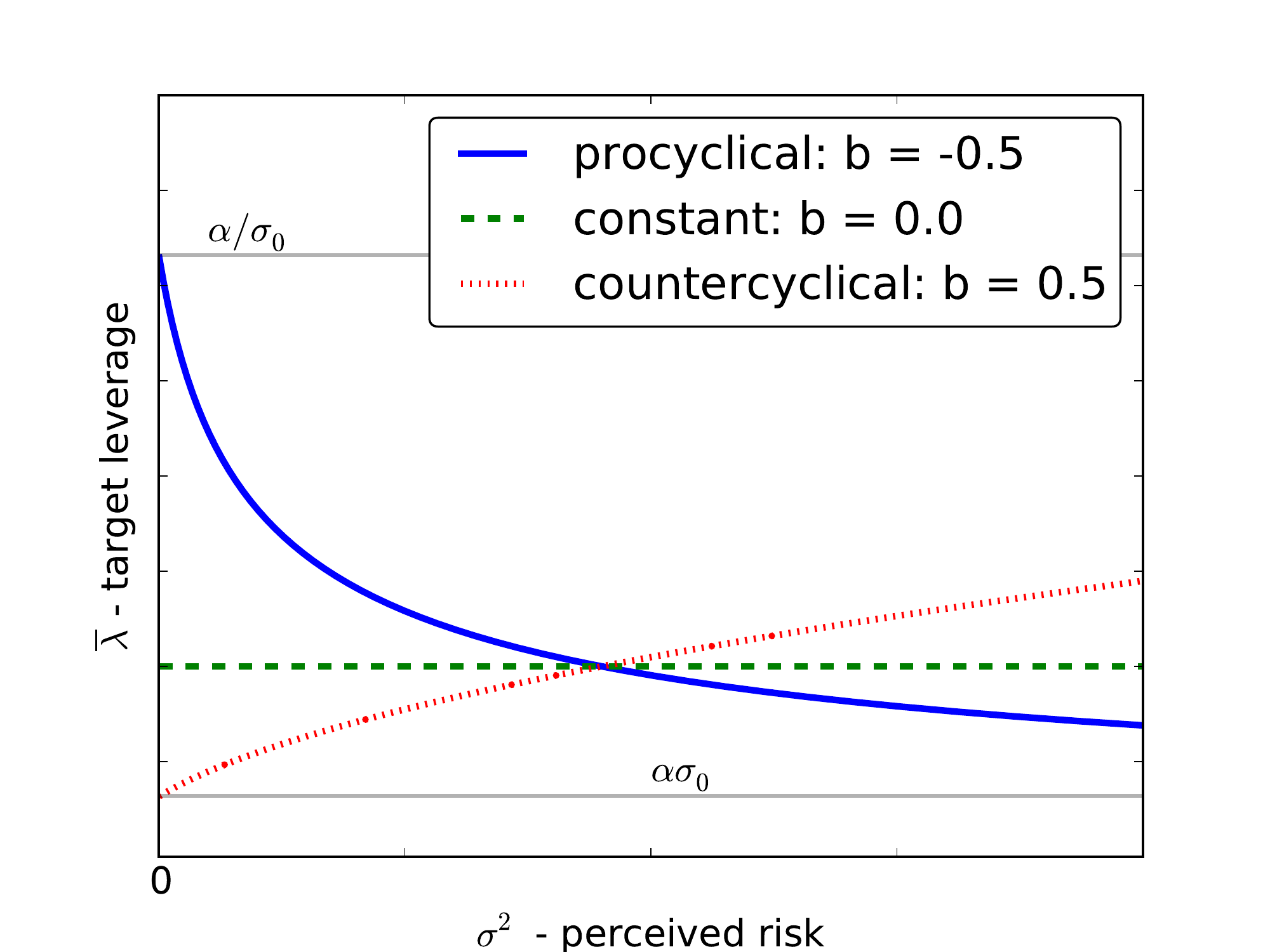}
\caption{Illustration of target leverage as a function of perceived risk based on Equation (\ref{EQ::leverage_policy}) with $\sigma_0^2 > 0$. Continuous blue line: procyclical policy with $b = -0.5$. Dashed green line: constant leverage policy with $b=0$. Dotted red line: countercyclical policy with $b=0.5$.  Continuous grey lines show relation to $\sigma_0$.}\label{FIG::intuition}
\end{figure}

The parameter $b$ is called the {\it cyclicality parameter}, due to the fact that $F_{(\alpha,\sigma^2_0,b)}$ is procyclical for $b<0$ and countercyclical for $b>0$ (see Definition \ref{Def:cyclicality}).  For procyclical policies the leverage is inversely related to risk, i.e. leverage is low when risk is high and vice versa. For countercyclical policies the opposite is true; when risk is high leverage is also high, see Figure \ref{FIG::intuition}.  It is important to note that our definition of policy cyclicality does not refer to macroeconomic measures such the credit-to-GDP ratio or asset prices. Instead it is defined solely by the bank's response to changes in perceived risk. In this sense the countercyclical policies proposed in this model differ from the countercyclical capital buffer proposed by the Bank of England, see \cite{FPC_2014}, which keys off the credit-to-GDP ratio.

Finally, the parameter $\sigma^2_0$ bounds the leverage control policy from above for $b < 0$ such that $F_{(\alpha,\sigma^2_0,b)} \leq \alpha ( \sigma_0^2 )^b$ and from below for $b>0$ such that $F_{(\alpha,\sigma^2_0,b)} \geq \alpha ( \sigma_0^2 )^b$.  For the remainder of this paper we will restrict our analysis to leverage control policies of the class $F_{(\alpha,\sigma^2_0,b)}$. We illustrate the three corner cases of the above class of leverage control policies with $\sigma_0^2 > 0$ in Figure \ref{FIG::intuition}.

\subsection{Asset price dynamics}

The bank's target leverage $\bar{\lambda}(t)$ at time $t$ defines a target portfolio value $\bar{A}_{\text{B}}(t) = \bar{\lambda}(t) E_{\text{B}}(t)$, where $E_{\text{B}}(t)$ is the equity of the bank. The difference between the target portfolio and the current portfolio then determines the change of the balance sheet $\Delta B(t)$ required for the bank to achieve its target leverage: 
\begin{itemize}
\item If $\Delta B(t) > 0$, the bank will borrow $\Delta B(t)$ and invest this amount into the risky and the risk free asset according the bank's portfolio weights. 
\item If $\Delta B(t) < 0$, the bank will liquidate part of its portfolio and pay back $\Delta B(t)$ of its liabilities.
\end{itemize}

The evolution of the fund's portfolio weight in the risky asset depends on the asset's price relative to a constant fundamental value $\mu$, and also on random innovations. The fund investor therefore combines two economic mechanisms:  (1) The constant fundamental value means that the price of the risky asset is ultimately anchored on the performance of unmodeled macro-economic conditions, which we assume are effectively constant over the length of one run of our model.
(2) We allow random innovations in the portfolio weight that reflect exogenous shocks.
  
We assume that the fund invests a fraction $w_{\text{F}}(t)$ of its total assets in the risky asset, and that the time evolution of $w_{\text{F}}(t)$ follows a mean reverting process with a GARCH(1,1) noise term. Thus, the fundamentalist investor provides a source of time varying exogenous volatility to the model. It is important that the exogenous volatility is time varying as this motivates the need for microprudential leverage control:  To minimize risk, the bank must estimate the expected future volatility and adjust its leverage accordingly.  

Given the aggregate demand of the bank and the fund, and assuming for simplicity that there is a supply of exactly one unit of the risky asset that is infinitely divisible, the price of the risky asset is determined through market clearing by equating demand and supply.

\subsection{Time evolution}

The model evolves in discrete time-steps of length $\tau$.  We make this a free parameter so that the model has well-defined dynamics in the continuum limit $\tau \to 0$, which is useful for calibration.  At each time-step the bank and the fund update their balance sheets as follows:
\begin{itemize}
\item The bank updates its historically-based estimate of future volatility and computes its new target leverage accordingly.  Volatility estimation is done using an exponential moving average with an approach similar to RiskMetrics (see \cite{Morgan1996});
\item The bank pays dividends or raises capital  to reach its target equity $\overline{E}$;
\item The bank determines how many shares of the risky asset it needs to trade to reach its target leverage;
\item At the same time, the fundamentalist fund submits its demand for the risky asset;
\item The market clearing price for the risky asset is computed and trades occur.\footnote{
It is important to note that the decision concerning equity and investment adjustments is taken before the current trading price of the risky asset is revealed.  We therefore assume that the bank uses the price of the previous time step as a proxy for the expected trading price, and acts accordingly. This assumption of myopic expectations marks a significant departure of our model from the general equilibrium setting of \cite{Adrian2012} and \cite{Adrian13}, but it is common in the literature on heterogeneous agents in economics (see for instance \cite{Hommes2006}).} 
\end{itemize}

\subsection{The model as a dynamical system}\label{SEC::6Dmap}
The dynamics of our model can be described as an iterated map for the state variable $x(t)$, defined as
\begin{equation}\label{EQ::var_vec}
x(t) = [ \sigma^2(t), w_{\text{F}}(t), p(t), n(t), L_{\text{B}}(t), p'(t) ]^\text{T},
\end{equation}
where $\sigma$ is the historical estimation of the volatility of the risky asset;  $w_{\text{F}}$ is the fraction of wealth invested by the fund in the risky asset; $p$ the current price of the risky asset; $n$ the share of the risky asset owned by the bank; $L_{\text{B}}$ the liabilities of the bank; and $p'$ is the lagged price of the asset, i.e. the price at the previous time step. 
 A detailed derivation of the model is presented in \ref{APP::Model}.  Here we simply present the model and provide some basic intuition.
Let us introduce the following definitions:
\begin{equation*}
\label{definitions}
\begin{aligned}
&\text{Bank assets}& A_{\text{B}}(t) &= p(t)n(t)/w_{\text{B}}, \\
&\text{Target leverage}& \bar{\lambda}(t) &= \alpha (\sigma^2(t) + \sigma^2_0)^b, \\
&\text{Balance sheet adjustment}& \Delta B(t) &= \tau \theta  (\bar{\lambda}(t) (A_{\text{B}}(t) - L_{\text{B}}(t)) - A_{\text{B}}(t) ),\\
&\text{Equity redistribution}& \kappa _{\text{B}}(t)&=  -\kappa _{\text{F}}(t) = \tau \eta  ( \overline{E} - (A_{\text{B}}(t) - L_{\text{B}}(t))),\\
&\text{Bank cash}& c_{\text{B}}(t) &= (1 - w_{\text{B}})n(t)p(t)/w_{\text{B}} + \kappa _{\text{B}}(t),\\
&\text{Fund cash}& c_{\text{F}}(t) &= (1 - w_{\text{F}}(t))(1 - n(t))p(t)/w_{\text{F}}(t) + \kappa _{\text{F}}(t).
\end{aligned}
\end{equation*}
The parameters $\theta$ and $\eta$ determine how aggressive the bank is in reaching its targets for leverage and equity respectively (i.e. the bank aims at reaching the targets on time horizons of the order $1/\theta$ and $1/\eta$).

The model can be written as a dynamical system in the form
\begin{equation}
\label{dynamicalSystem}
x(t+\tau) = g(x(t)),
\end{equation}
where the function $g$ is the following 6-dimensional map:
\begin{subequations}\label{EQ::6Dmap}
\begin{align}
\sigma^2(t+\tau) & =  (1- \tau \delta ) \sigma^2(t) + \tau \delta  \left(\log\left[\frac{p(t)}{p'(t)}\right] \frac{t_{\text{VaR}}}{\tau} \right)^2 ,\\
w_{\text{F}}(t+\tau) & =  w_{\text{F}}(t) +  \frac{w_{\text{F}}(t)}{p(t)} \left[\tau \rho(\mu - p(t))+\sqrt{\tau} s \xi(t) \right],\\
p(t+\tau) & =  \frac{w_{\text{B}}(c_{\text{B}}(t) + \Delta B(t)) + w_{\text{F}}(t+\tau)c_{\text{F}}(t)}{1 - w_{\text{B}} n(t) - (1 - n(t))w_{\text{F}}(t+\tau)},\\
n(t+\tau) & =  \frac{w_{\text{B}}(n(t)p(t+\tau) + c_{\text{B}}(t) + \Delta B(t))}{p(t+\tau)},\\
L_{\text{B}}(t+\tau) & =  L_{\text{B}}(t) + \Delta B(t), \\
p'(t+\tau) & = p(t).
\end{align}
\end{subequations}
Each of these equations can be understood as follows:
\begin{enumerate}[(a)]
\item The expected volatility $\sigma^2$ of the risky asset is updated through an exponential moving average. The parameter $\tau \delta\in(0,1)$ defines the length of the time-window over which the historical estimation is performed, while the parameter $t_\text{{VaR}}$ represents the time-horizon used by the bank in the calculation of VaR.
\item
The adjustment of the fund's risky asset portfolio weight $w_{\text{F}}$ drives the price towards the fundamental value $\mu$, with an adjustment rate $\tau \rho\in(0,1)$.  The demand of the fund also depends on exogenous noise, which is assumed to be a normal random variable $\xi(t)$ with amplitude $s(t) \ge 0$.  The amplitude varies in time so that the variable $\chi(t) = s(t) \xi(t)$ follows a GARCH(1,1) process.  The factors of $\tau$ guarantee the correct scaling as $\tau \to 0$.
\item The market clears.   $c_{\text{F}}(t)$ and $c_{\text{B}}(t)$ are the amount of cash held respectively by the fund and the bank.
\item The bank ownership of the risky asset $n(t + 1)$ adjusts according to market clearing.  
\item Bank liabilities are updated to account for the change $\Delta B(t)$ in the asset side of the balance sheet. 
\item
The lagged price variable $p'(t)$ is required to complete the state vector, and make the map a first order dynamical system of the form given in Equation (\ref{dynamicalSystem}). 
\end{enumerate}
\section{Examples of leverage cycles}\label{SEC::dynamcis}
In order to explore the dynamical behavior of the model we solve it numerically. For now we will only consider leverage control policies that are procyclical; in particular we choose $b=-0.5$ throughout this section, corresponding to the case of risk management under VaR. 

\begin{table}
\begin{tabular}{lllrl} 
\toprule
& Symbol & Description & Default & Unit \\
\midrule
Bank & $\tau$ & Time step & $0.1$ & year \\
	 & $\delta$ & Memory for volatility estimation & $0.5$ & $\text{year}^{-1}$ \\ 
     & $t_{\text{VaR}}$ & Horizon for VaR calculation & $0.1$ & year \\ 
	 & $\sigma^2_0$ & Risk offset & $10^{-6}$ & 1 \\ 
	 & $b$ & Cyclicality of leverage control & $-0.5$ (v) & 1 \\
	 & $\alpha$ & Risk level & 0.075 (v) & 1 \\
	 & $\overline{E}$ & Bank's equity target & $2.27$ (v) & 1 \\ 
	 & $w_{\text{B}}$ & Bank's weight for risky asset & $0.3$ (v) & 1 \\ 
	 & $\theta$ & Balance sheet adjustment speed & $9.5$ (v) & $\text{year}^{-1}$ \\ 
	 & $\eta$ & Equity redistribution speed & $10$ & $\text{year}^{-1}$ \\
\midrule
Fund & $\mu$ & Fundamental value & $25$ & 1 \\ 		
	 & $\rho$ & Mean reversion & $0.1$ & $\text{year}^{-1}$\\ 	
\midrule 
GARCH & $a_0$ & Baseline return variance & $10^{-3}$ & 1 \\
	  & $a_1$ & Error autoregressive term & $0.016$ & 1 \\
	  & $b_1$ & Variance autoregressive term & $0.87$ & 1 \\

\bottomrule 
\end{tabular} 
\caption[Parameters of extended variable equity model]{Overview of parameters for the numerical model solution. Values marked with (v) indicate that they are subject to change from their default values; ``1" indicates that the parameter is dimensionless.}
\label{TAB::param_overview}
\end{table}

\subsection{Model calibration}
While this model is too stylized to be fully calibrated, approximate values for some key parameters can be obtained. This then allows us to test the realism of some of the properties of the model. In the following we will briefly discuss the choice and effect of these key parameters, including the timescale of the risk estimation, the balance sheet adjustment speeds and bank riskiness. A full list of parameters is provided in Table \ref{TAB::param_overview}. 

\subsubsection*{Timescale parameters}

We have carefully constructed the dynamical system so that it reaches a continuum limit as $\tau \to 0$. For computational efficiency we choose $\tau$ to be the largest possible value with behavior similar to that in the continuum limit, which results in a time step of $\tau = 0.1$ years.  As long as $\tau$ is this size or smaller the results change very little.

The parameter $\delta$ sets the timescale for the exponential moving average used to estimate volatility, and is the most important determinant of the overall timescale of the dynamics.  The characteristic time for the moving average is $t_\delta = 1/\delta$.\footnote{
The contribution to the moving average of a squared return $y(t)$ observed at time $t$ is $y(t+\Delta t) = (1-\tau \delta)^{\Delta t/\tau} y(t)$ at time $t+\Delta t$. We define the typical time $t_{\delta}$ such that $y(t+t_\delta)/y(t) = 1/e$. Thus $t_\delta = -\tau/\log[1-\tau\delta] \approx 1/{\delta}$ for $\tau\delta \ll 1$.} 
According to the RiskMetrics approach \cite{Morgan1996}, the typical timescale used by market practitioners is $t_{\delta} \approx 2 \text{ years}$.  We thus have the luxury of being able to calibrate this parameter from ``first principles".  We therefore set $\delta = 0.5 \text{ year}^{-1}$, corresponding to a two year timescale, and keep it fixed throughout.

Another timescale parameter is $t_{\rm VaR}$, the time horizon over which returns are computed for regulatory purposes. In practice, the timescale for the regulatory capital requirements varies depending on the liquidity of the asset portfolio and ranges from days to years.  A good rule of thumb is to choose $t_{\rm VaR}$ roughly equal to the time needed to unwind the portfolio.  We assume $t_{\rm VaR} = \tau = 0.1 \text{ years}$, i.e. a little more than a month.

The parameters $\theta$ and $\eta$ define how aggressive the bank is in reaching its target for leverage and equity.  Our default assumption is that the bank tries to meet its target on a timescale of about one time step of the dynamics, and so unless otherwise stated, in the following we set $\theta=9.5 \text{ year}^{-1}$ and $\eta=10 \text{ year}^{-1}$. This ensures that the bank's realized leverage is always close to its target. We will vary the parameter $\theta$ and discuss how it affects the stability of the dynamics in Section \ref{SEC::adjustment_speed}. 

\subsubsection*{Market power of the bank}

The dynamics of this model depend on the competition between the stabilizing properties of the fundamentalist and the destabilizing properties of the bank.  Thus to understand the parameters and their effect on the dynamics is it useful to understand how they influence the market power of the bank, which is roughly speaking the product of the leverage $\lambda$ and the relative size of the banking sector $R$.  To get a feeling of this, we show in Equation \eqref{EQ::rel_size} in the appendix that at the fixed point equilibrium the parameters $\overline{E}$, $w_\text{B}$, $w_\text{F}$, $\mu$, $\sigma_0$, and $\alpha$ all jointly determine the fraction of the risky asset $R$ owned by the bank. Note that the numerical values chosen for the target equity $\overline{E}$ and the fundamental price $\mu$ are arbitrary -- only their ratio is important. We choose the values of the above parameters in order to produce the desired value of $R$ (though we often vary $\alpha$ independently).  Note that the bank represents all investors with leverage targets,  and the set of institutions with a comparable Value-at-Risk based leverage constraint\footnote{In principle the constraint can be either imposed by a regulator, creditors or internal risk management.} is larger than the banking sector.   

The other key parameter affecting the stability of the model is the bank riskiness $\alpha$.  Increasing $\alpha$ increases both the bank's market power and its default risk.   Note that $\alpha$ is also related to $t_{\rm VaR}$ by the fact that, all else equal, increasing the timescale over which the risk is measured corresponds to taking more risk.  (Increasing $\alpha$ usually increases leverage, though as discussed in footnote \ref{alphaFootnote} this is not always true).  The bank's portfolio weight $w_{\text{B}}$ for the risky asset has a similar effect to the bank's equity target $\overline{E}$.\footnote{
In fact, it can be shown the the critical leverage is inversely proportional to $w_{\text{B}}$.}
Increasing any of these parameters increases the bank's market power.  

In our calibration we choose a particular level of bank riskiness $\alpha$ and the relative size of the bank and the fund to match two basic properties of the run up and the subsequent collapse of leverage and asset prices during global financial crisis in 2008/2009. First, we seek a peak to trough ratio in the price of the risky asset of roughly $2$. Second, we target a period of oscillation of roughly ten years. Matching these calibration targets comes at the price of achieving realistic levels of bank leverage in our simulations. Given our choice for $\alpha$ we obtain levels of bank leverage of around $6$. This is below typically observed levels of leverage of around $20$. The fact that we cannot calibrate our model to match several calibration targets simultaneously is a clear weakness of the model, but is not surprising given its simplicity.  It should also be noted that due to hedging banks may be able to achieve levels of risk that are much lower than that of a single bare asset as we model here, and this may explain the discrepancy in leverage.

Finally, we pick parameters for the fund GARCH process $a_0$, $a_1$ and $b_1$ in order to achieve a randomly perturbed asset price path that still follows a leverage cycle roughly as observed in Figure \ref{FIG::data_leverage_cycle}.

\subsection{Overview of model dynamics}
We now build some intuition about the model dynamics. First, consider the extreme case where $\overline{E} \rightarrow 0$, i.e. where the market power of the bank is negligible so that the price dynamics are dominated by the fund. This is the purely \textit{microprudential} case where the bank's actions have no significant effect on the market and the only source of volatility is exogenous.   In this case ($s>0$) we expect the price to perform a mean reverting random walk around the fundamental price $\mu$. In the deterministic case, i.e. $s=0$, the fund updates its portfolio weight until $p(t) = \mu$, i.e. until the price has converged to the fundamental price and the system settles to a fixed point equilibrium.

When $\overline{E}$ is large enough that the bank has a significant impact on the price process the dynamics are less straightforward. We refer to this scenario as the \textit{macroprudential} case. Suppose, for example, that there is a negative shock in the investment of the fund. This negative shock will lead to an increase in the perceived risk $\sigma^2(t)$. Under a procyclical leverage control policy an increase in perceived risk causes a decline in the bank's leverage constraint. As a consequence the bank will have to deleverage in the time step following the negative shock, i.e. $\Delta B(t) < 0$. If the bank decreases its position and it has non-negligible market impact, the price will drop for $\Delta B(t) < 0$ ceteris paribus. This is clear from Equation \ref{EQ::clearing_price} in the appendix. Thus an initial negative shock can be amplified by the bank's deleveraging response. This destabilizing feedback loop is a key ingredient for what is to come and distinguishes risk management in the macroprudential case from the microprudential case. In the macroprudential case the bank's risk management affects the system's state and introduces endogenous volatility on top of exogenous volatility.

\begin{figure}
\centering
\includegraphics[width =\textwidth]{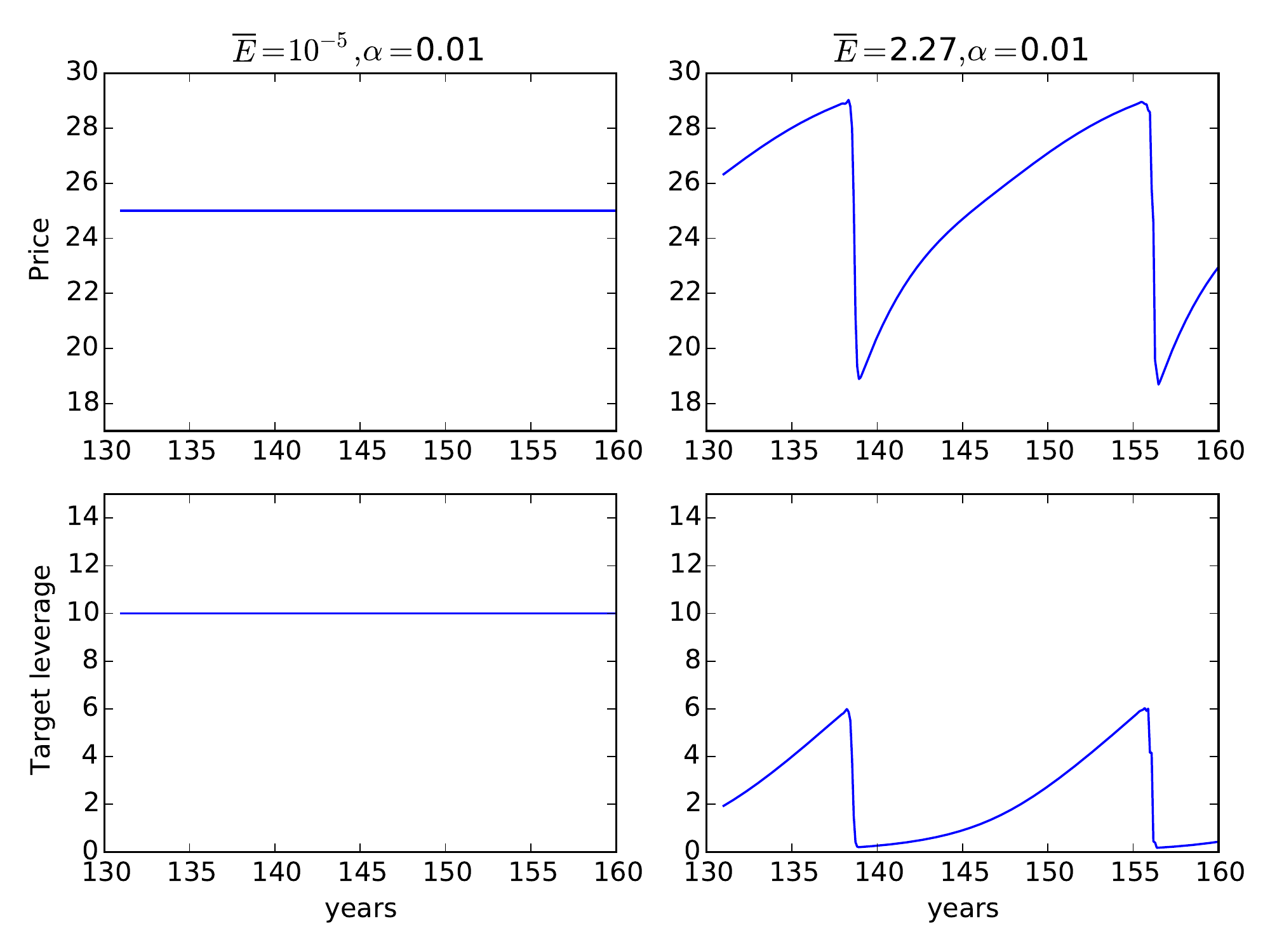}
\caption{Time series of price and leverage in the deterministic case. Left panel: scenario (i) -- microprudential, the fund dominates the bank ($\overline{E} = 10^{-5}$), i.e. the bank has no significant market impact. In this case the system goes to a fixed point equilibrium where the leverage and price of the risky asset remain constant.  Right panel:  scenario (i)  -- macroprudential, the bank has significant market impact ($\overline{E} = 2.27$). In this case the bank's risk management leads to persistent oscillations in leverage and price of the risky asset with a time period of roughly 15 years.}\label{FIG::example1_2}
\end{figure}

\begin{figure}
\centering
\includegraphics[width =\textwidth]{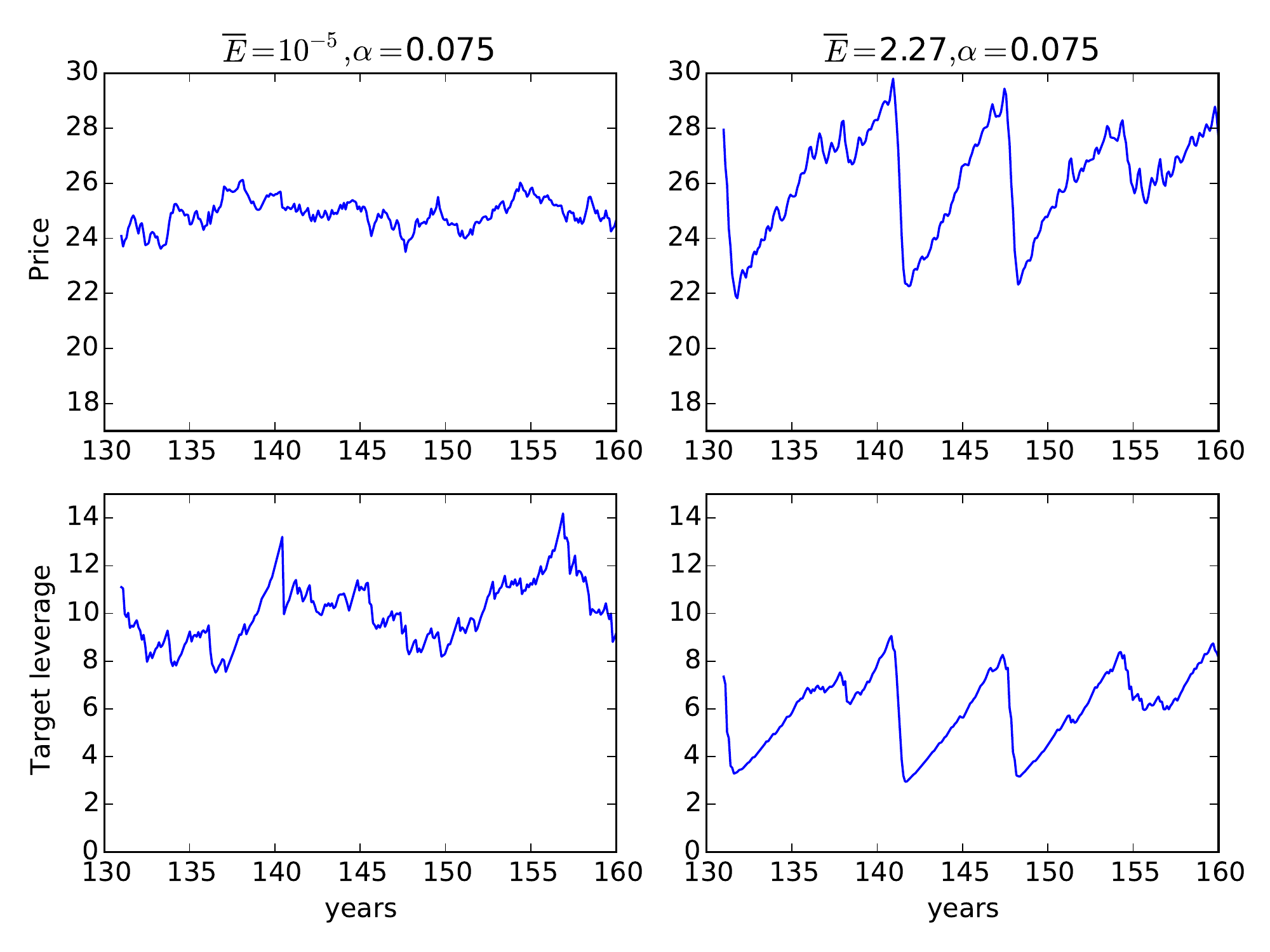}
\caption{Time series of price and leverage in the stochastic case. Left panel: scenario (iii) -- microprudential, the fund dominates the bank ($\overline{E} = 10^{-5}$), i.e. the bank has no significant market impact. In this case the price is driven by the fund's trading activity and performs a mean reverting random walk around the fundamental value $\mu = 25$.
Right panel: scenario (iv) -- macroprudential, the bank has significant market impact ($\overline{E} = 2.27$). In this case the bank's risk management leads to irregular oscillations in leverage and price of the risky asset that are similar to the deterministic case.}\label{FIG::example3_4}
\end{figure}

To illustrate the dynamics of the model we will investigate the following four scenarios:
\begin{enumerate}[(i)]
\item Deterministic, microprudential: $\overline{E} = 10^{-5}$ and $s=0$. 
\item Deterministic, macroprudential: $\overline{E} = 2.27$  and $s=0$.
\item Stochastic, microprudential:  $\overline{E} = 10^{-5}$ and $s>0$.
\item Stochastic, macroprudential: $\overline{E} = 2.27$ and $s>0$.
 \end{enumerate}
Unless otherwise stated all parameters are as specified in Table \ref{TAB::param_overview}.   The first two cases are for the deterministic limit with $s = 0$, which is useful to gain intuition.  The last two cases are with more realistic levels of exogenous noise.  We summarize our results for scenarios (i) and (ii) in Figure \ref{FIG::example1_2} and for scenarios (iii) and (iv) in Figure \ref{FIG::example3_4}. 

The microprudential scenarios (i) and (iii) behave as expected:  In the deterministic limit the systemic simply settles into a fixed point with prices equal to fundamental values.  When there is exogenous noise the system makes excursions away from the fixed point but never drifts far away from it, and the dynamics remain relatively simple.

In contrast the macroprudential scenarios (ii) and (iv) display large oscillations both in leverage and price.  We refer to this oscillation as the {\it Basel leverage cycle}.  Surprisingly, the oscillations occur even in the deterministic limit, i.e. without any external shocks. During the cycle the price and leverage slowly rise and then suddenly fall, with a period of about $\Delta t \approx 15  \text{ years}$ in the deterministic case.

In the stochastic case we observe a period of about $\Delta t \approx 10  \text{ years}$. This is roughly on the order of magnitude of the period of the Great Moderation and the subsequent financial crisis. Note that the period of oscillation depends strongly on the risk estimation horizon $t_{\delta}$, but this is set to two years based on behavioral data.\footnote{
In fact, the period is roughly proportional to $1/t_{\delta}$. The period becomes large for very low values of $\delta \tau$ and then declines as the risk estimation horizon is increased -- for the range of $\tau \delta$ the variation of the period ranges over roughly two orders of magnitude. The period also depends on the values of $\eta$, $t_{\text{VaR}}$, $\theta$, $w_B$ and $\alpha$. As $\eta$ or $t_{\text{VaR}}$ are increased the period increases. As $\theta$, $\alpha$ and $w_B$ are increased, i.e. as the system moves to a more unstable regime, the period declines. However, for these parameters the variation in the period over the parameter range is only on roughly one order of magnitude. The period of oscillation is robust to changes in $w_{\text{F}}$.
}

The oscillations have the following economic interpretation: Suppose we begin at about $t = 140 \text{ years}$ in the left panel of Figure \ref{FIG::example1_2},  with leverage low, perceived risk high, and prices low but increasing.  The perceived risk slowly decreases as the memory of the past crisis fades.  From a mechanical point of view this is due to the smoothing action of the exponential moving average.    As the moving average is updated on each timestep, the volatility $\sigma^2$ decreases; this causes the leverage to increase, and the bank buys more shares to meet its increased leverage target.   The change in price is lower than the current historical average, so on the next step the volatility $\sigma^2$ drops, driving the leverage higher.   As the leverage becomes very high the system becomes increasingly fragile.  In the phase space the system approaches a hyperbolic fixed point where the leverage is so large that that a crash occurs.  The downward crash ultimately comes to an end by the increasingly heavy investment of the fundamentalist fund.  After the crash volatility is high and leverage is low, and the cycle repeats itself.  

The fragility that drives the crashes comes from the fact that at high levels of leverage a small increase in risk is sufficient to cause a drastic tightening of the leverage constraint.   This intuition can be made precise by comparing the derivative of the leverage control policy for high vs. low leverage; for convenience we take $\sigma^2_0 \ll 1$.\footnote{
For $b < 0$ the parameter $\sigma_0$ imposes a cap on the target leverage; larger values for $\sigma_0^2$ would make this unrealistically low.}
The result is that
\begin{equation*}
\frac{dF_{(\alpha,\sigma^2_0,-0.5)}}{d\sigma^2(t)}(\sigma^2(t)) = \begin{cases}
-0.5 /\sigma_0 \ll 0 &\text{, for } \sigma^2(t) \rightarrow 0 \ \land \ \sigma^2_0 \ll 1 \\
 0 &\text{, for } \sigma^2(t) \rightarrow \infty \ \land \ \sigma^2_0 \ll 1 \\
\end{cases}
\end{equation*}
In the high leverage limit, i.e. when the perceived risk is small, the sensitivity of the leverage target $F$ to variations in risk tends to infinity.  In contrast the sensitivity is zero in the opposite limit where leverage is low and perceived risk is large. 
Thus increasing leverage of the banking system has a two-fold destabilizing effect:  It can make the dynamics unstable and lead to chaos, but it also makes it more sensitive to shocks, which can result in sudden deleveraging.  

\begin{figure}
\centering$
\begin{array}{cc}
\includegraphics[width =0.5\textwidth]{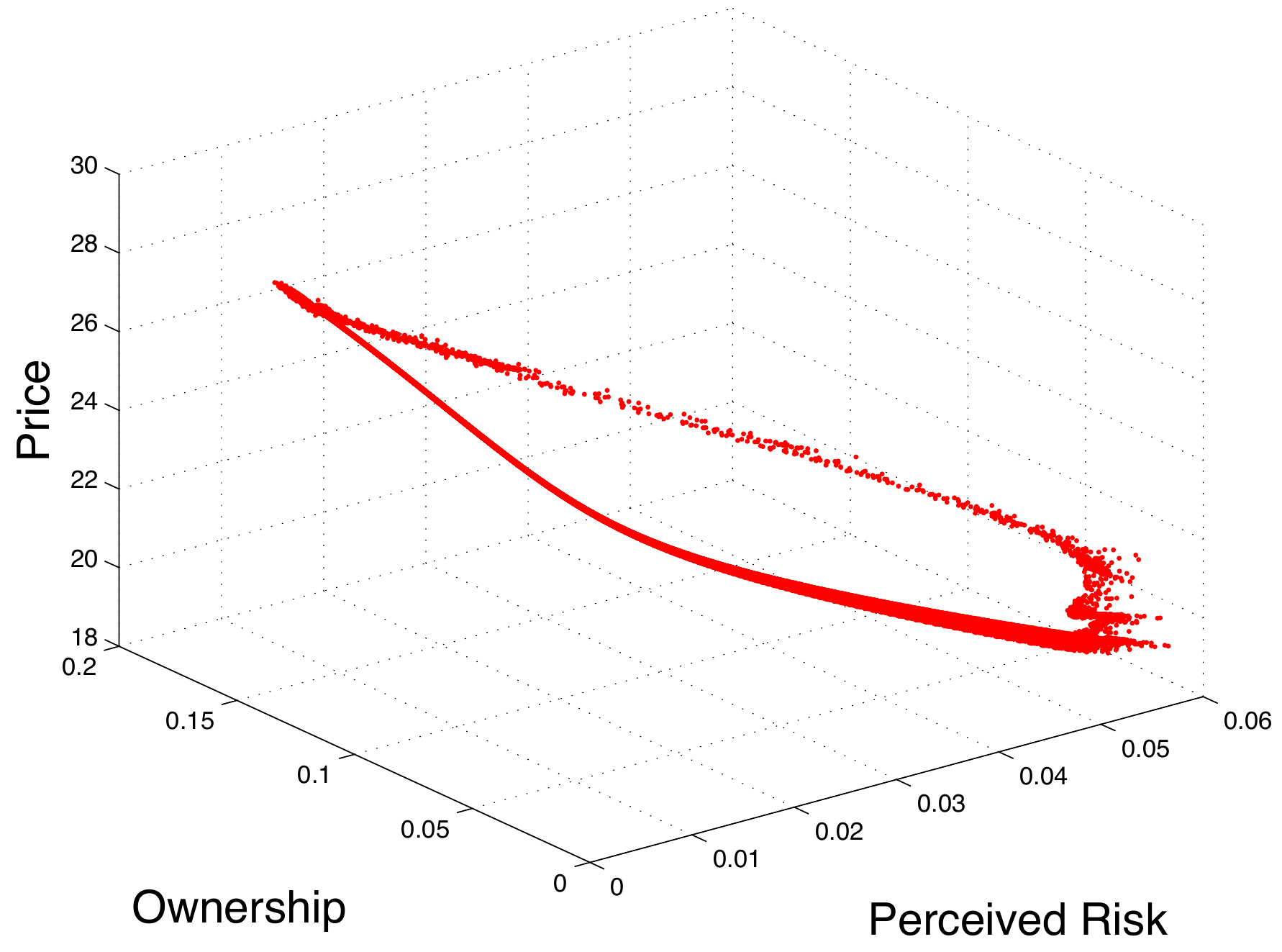}
\includegraphics[width =0.4\textwidth]{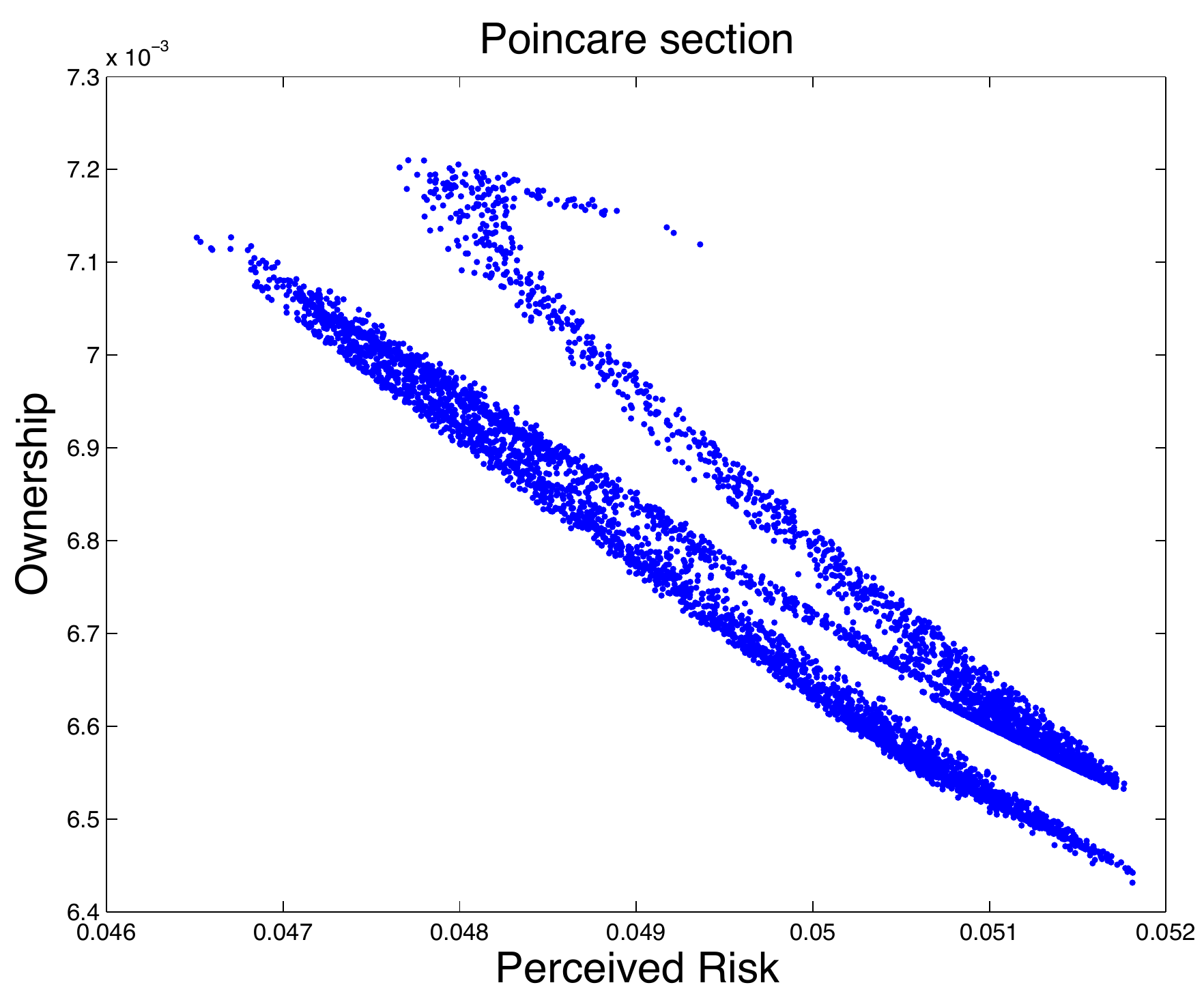}
\end{array}$
\caption{Left panel: Three dimensional phase plot of the system's attractor when there is a deterministic leverage cycle. Right panel: A Poincar\'e section is constructed by recording values for the bank ownership of the risky asset (y-axis) and the perceived risk (x-axis) whenever the price is increasing and $p(t) = 20$, repeating for $10^6$ time steps.  This exhibits the characteristic stretching and folding associated with chaotic dynamics.}\label{FIG::poincare}
\end{figure}

The leverage cycles are not strictly periodic due to the fact that the oscillations are chaotic.  This becomes clearer by plotting the dynamics in phase space and then taking a Poincar\'e section, as illustrated in Figure \ref{FIG::poincare}.  The phase plot makes the cyclical structure clearer; the 3D representation shows how the ownership of the risky asset varies during the course of the leverage cycle (left panel). The Poincar\'e section is constructed by plotting ownership vs. perceived risk every time the trajectory crosses the hyper-plane $p(t)=20$ with the price increasing (right panel). The Poincar\'e section shows the characteristic fractal structure, and shows the stretching and folding that makes the dynamics chaotic.  The fact that these dynamics are chaotic is confirmed in the next section, where we do a stability analysis and compute the Lyapunov exponent.

In summary, depending on the choice of parameters, the model either goes to a fixed point (scenario (ii)) or shows chaotic irregular cycles (scenarios (i) and (iii)).  As expected the dynamics become more complicated when noise is added, but the essence of the Basel leverage cycle persists even in the zero noise limit.

\section{Determinants of model stability}

\subsection{Deterministic case}

In the deterministic case the standard tools of linear stability analysis can be used to characterize the boundary between the fixed point equilibrium and leverage cycles.  In this section we will use this to characterize the behavior of the system as the risk parameter $\alpha$ and the cyclicality parameter $b$ are varied.  We begin by studying the deterministic case, where we can compute things analytically, and then present numerical results for the stochastic case.   The details of the stability analysis are presented in \ref{APP::Model}.   

The system has a unique fixed point equilibrium $x^*$, given by
\begin{equation}\label{EQ::fixed_point}
\begin{aligned}
x^* &= (\sigma^{2*},w_{\text{F}}^*,p^*,n^*,L_{\text{B}}^*,p'^*) \\
	&= (0,w_{\text{F}}(0),\mu, \frac{1}{\mu} \alpha\sigma^{2b}_0\overline{E}w_{\text{B}}, (\alpha\sigma^{2b}_0 - 1)\overline{E},\mu).
\end{aligned}
\end{equation}
This corresponds to a leverage $\lambda^*$ and relative size of bank to fund $R_c(x^*)$, given by
\begin{equation}
\begin{aligned}
\lambda^* &= \alpha \sigma_0^{2b},\\
R(x^*) &= \frac{A_{\text{B}}^*}{A_{\text{F}}^*} = \frac{\lambda^* E_{\text{B}}^*}{(1 - n^*)p^*/w_{\text{F}}^*}.
\end{aligned}\label{EQ::crit_lev}
\end{equation}
At the equilibrium $x^*$ the price is constant at its fundamental value and the bank is at its target leverage. The stability of the equilibrium depends on the parameters.  Regime (i) observed in the numerical simulations of the previous section corresponds to the stable case.  In this case, regardless of initial conditions, the system will asymptotically settle into the fixed point $x^*$.  In contrast, when the fixed point $x^*$ is unstable there are two possibilities.  One is that there is a leverage cycle, in which the dynamics are locally unstable but exist on a chaotic attractor that is globally stable; the other is that the system is globally unstable, in which case the price either becomes infinite or goes to zero. 

\begin{figure}
\centering$
\begin{array}{cc}
\includegraphics[width=0.70\textwidth]{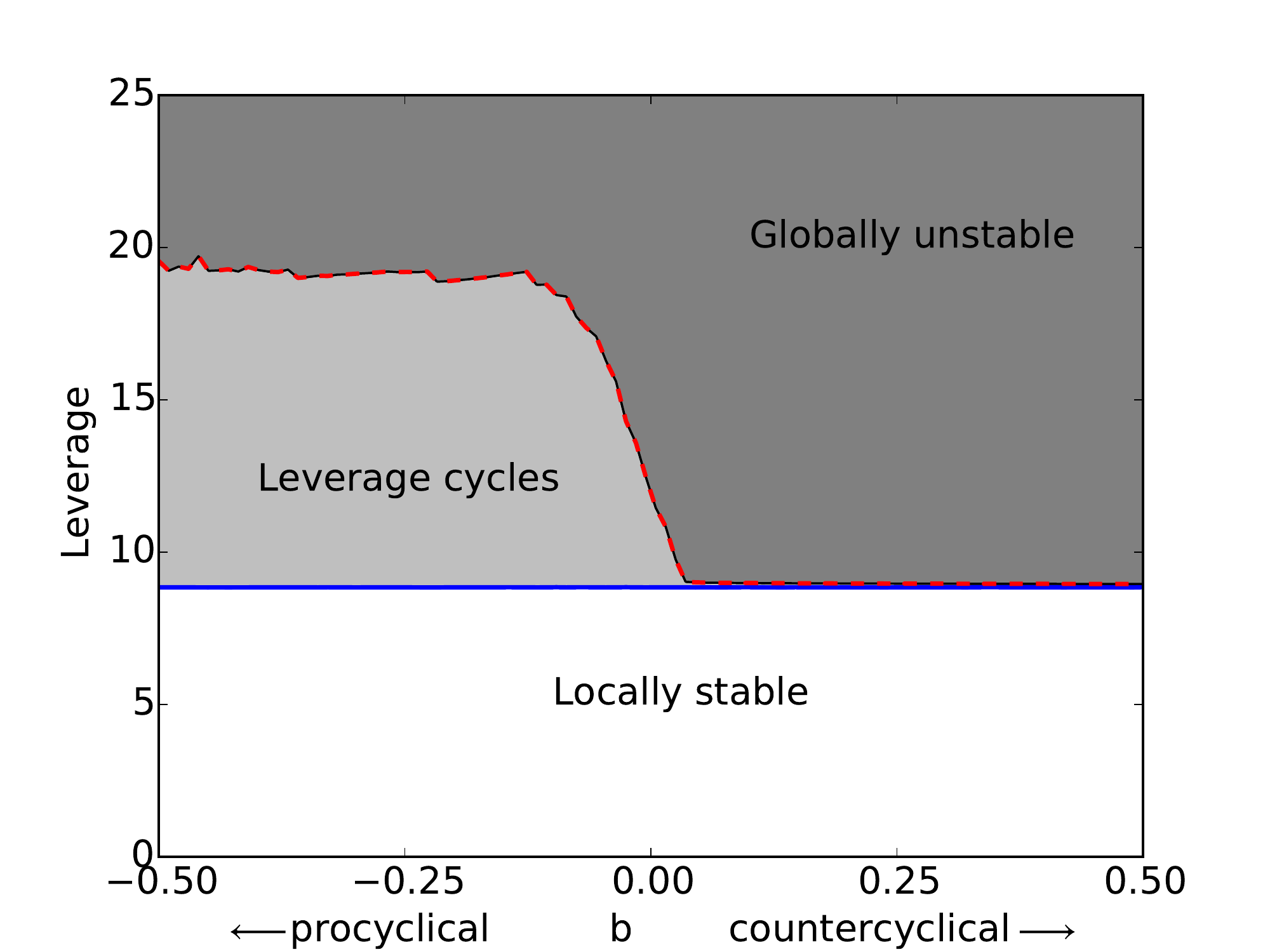}
\end{array}$
 \caption{A bifurcation diagram showing the three regimes in the deterministic case.  The risk parameter $\alpha$ and the cyclicality parameter $b$ are varied while holding the other parameters constant at the value in Table 1.  The white region corresponds to a stable fixed point equilibrium, the light gray region to leverage cycles and the dark gray region to global instability.  The blue line corresponds to the critical leverage $\lambda^*_c$ in Equation \ref{EQ::crit_lev} at the critical value $\alpha_c$ where the fixed point becomes unstable.  }\label{FIG::stability}
\end{figure}

In Figure \ref{FIG::stability} we show the results of varying the risk parameter $\alpha$ and the cyclicality parameter $b$.   The risk parameter $\alpha$ provides the natural way to vary the risk of the bank, but the realized risk for a given $\alpha$ depends on parameters due to other factors such as changes in volatility.   For diagnostic purposes leverage is a better measure.\footnote{
While $\alpha$ tends to increase leverage, when the leverage control policy is procyclical the behavior is not always monotonic.  This is because increasing $\alpha$ tends to increase volatility, but increasing volatility drives the target leverage down, so the two effects compete with each other.\label{alphaFootnote}}
Figure \ref{FIG::stability} shows each of the three regimes, corresponding to the stable equilibrium, leverage cycles or global instability, as a function of the leverage and the cyclicality parameter $b$.  The boundary where the fixed point equilibrium becomes unstable is computed analytically based on the leverage $\lambda^*_c$ where the modulus of the leading eigenvalue is one.  The boundary for globally unstable behavior is more difficult to compute as it requires numerical simulation.

This diagram reveals several interesting results.  As expected, for low leverage the system is stable and for higher leverage it is unstable.  Somewhat surprisingly, the critical leverage $\lambda^*_c$ is independent of $b$, and consequently the size of the regime with the stable equilibrium is unaffected by whether the leverage control is procyclical or countercyclical.  In the procyclical regime there is a substantial area of parameter space with leverage cycles.  For the countercyclical regime, in contrast, there is only a small regime with leverage cycles.  Throughout most of the parameter range the system makes a direct transition from the stable fixed point equilibrium to global instability.  The instability is not surprising:  In the countercyclical regime there is an unstable feedback loop in which increasing leverage drives increasing prices and increasing volatility, which further increases the leverage.  Thus for high leverage there are unstable regimes for both pro- and countercyclical behavior, but the instability is even worse in the countercyclical regime.

\subsection{Stability when there is exogenous noise}

In the case where there is exogenous noise we can only measure the stability numerically.  This is done by computing the largest Lyapunov exponent of the dynamics.  The Lyapunov exponents are a generalization of eigenvalues that apply to trajectories that are more complicated than fixed points.   The leading Lyapunov exponent measures the average rate at which the separation between two nearby points changes in time -- when the dynamics are locally stable nearby points converge exponentially and the leading Lyapunov exponent is negative, and when they are locally unstable nearby points diverge exponentially and the leading Lyapunov exponent is positive.  The Lyapunov exponent is a property of a trajectory, but for dissipative systems such as ours, it is also a property of the attractor.  A negative Lyapunov exponent implies a fixed point, and a positive Lyapunov exponent implies a chaotic attractor.   As expected, in the deterministic case we observe that leverage cycles have a positive leading Lyapunov exponent, confirming that the dynamics are chaotic.

It is also possible to compute Lyapunov exponents for stochastic dynamics.  To understand the basic idea of how this is done, imagine two realizations of the dynamics with the same sequence of random shocks, but starting at slightly different initial conditions, see \cite{Crutchfield198245}. Because the random noise is the same in both cases, it is possible to follow two infinitesimally separated points and measure the rate at which they separate.  If the leading Lyapunov exponent is positive this means that the dynamics will strongly amplify the noise.

\begin{figure}
\centering$
\begin{array}{cc}
\includegraphics[width=0.65\textwidth]{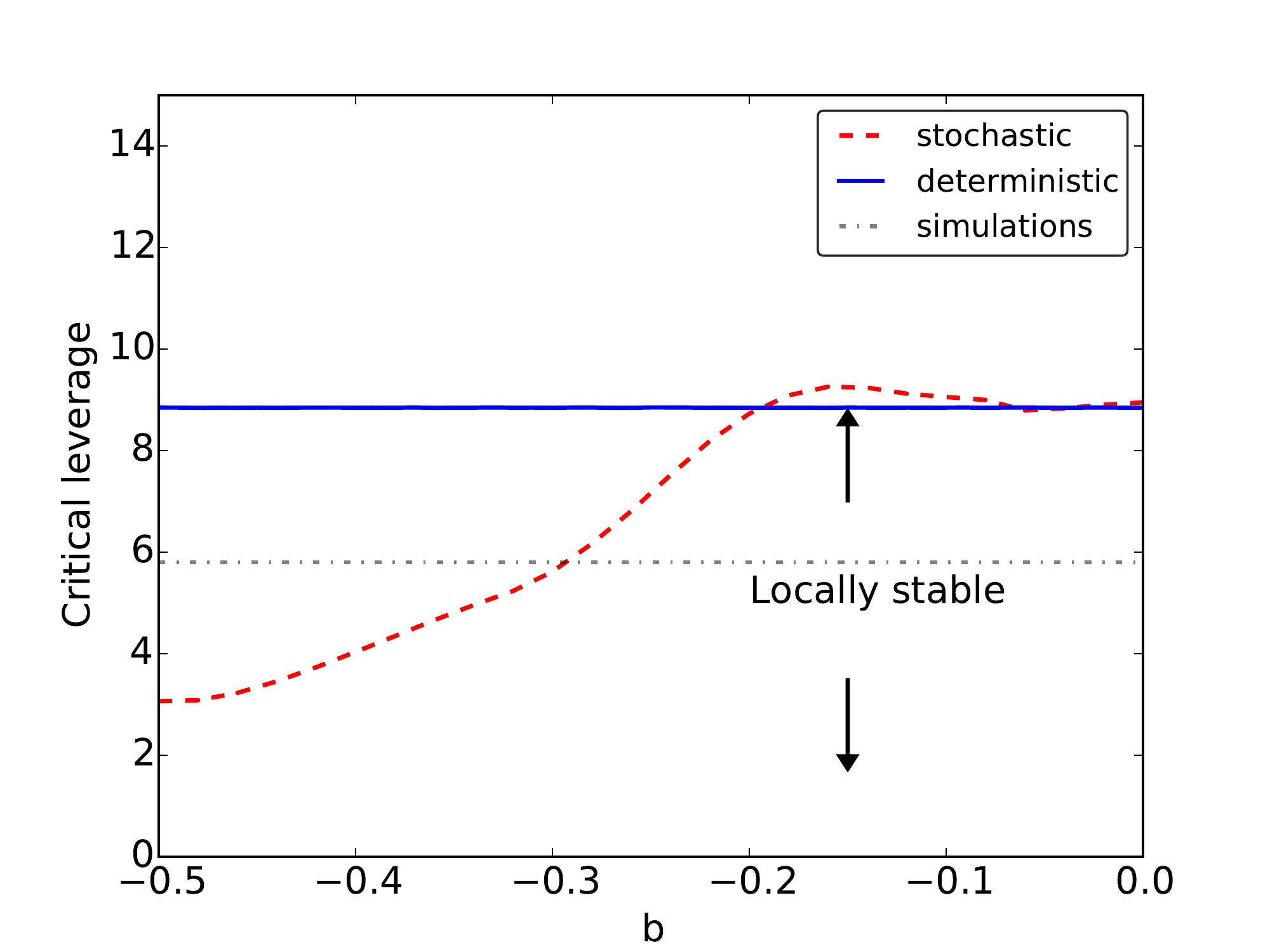}
\end{array}$
 \caption{A comparison of stability when the dynamics are deterministic vs. stochastic for the procyclical region ($b<0$). As in the previous figure, the critical leverage $\lambda^*_c$ for the deterministic case is shown as a blue line.  The dashed red line shows the parameter value where the dynamics become unstable as measured by the leading Lyapunov exponent; note the transition to chaos occurs at a much lower leverage.   The gray line shows the average target leverage obtained in the simulation, which is roughly independent of $b$.}\label{FIG::stochasticStability}
\end{figure}

We compare the stability for the stochastic and deterministic cases in Figure \ref{FIG::stochasticStability}.  This is done for the procyclical case only, since the direct transition from a fixed point to global instability in the countercyclical case complicates numerical work (and the countercyclical case is less relevant).
In the stochastic case the critical leverage is computed as the time average of the target leverage when the Lyapunov exponent becomes positive. Interestingly, the critical leverage in the stochastic case first starts below the deterministic critical leverage and then approaches it as $b$ is increased. This indicates that for strongly procyclical leverage control policies noise destabilizes the system. Somewhat surprisingly, the average leverage observed in the simulations is independent of $b$.

The most interesting conclusion from comparing the stochastic and deterministic cases is that when the dynamics are strongly procyclical (i.e. for $-0.5 < b < -0.2$) the noise significantly lowers the stability threshold.   In contrast, for larger values of $b > -0.2$ there is little difference in the stability threshold in the two cases.  This indicates that the dynamics becomes more stable when the leverage control policy is close to constant leverage.  This, together with the fact that in the countercyclical regime the system goes straight from stability to global instability, suggests that intermediate values of cyclicality (nearer to constant leverage) are likely to be most stable.

\subsection{Slower adjustment leads to greater stability}\label{SEC::adjustment_speed}

The bank's balance sheet adjustment speed $\theta$ has a strong effect on stability with interesting regulatory implications.  Intuitively, decreasing the adjustment speed should make the system more stable.   To take an extreme case, in the limit $\tau \theta \rightarrow 0$, the bank would hold its balance sheet constant regardless of changes in perceived risk. This would eliminate the feedback loop between asset prices, perceived risk and investment. Even when $\theta > 0$, decreasing the adjustment speed should have a stabilizing effect.\footnote{We have considered the case where the bank increases its leverage quicker than it decreases it. We have done this introducing an asymmetry in the parameter $\theta$ that controls the speed of leverage adjustement, i.e. introducing a parameter $\theta_+$ for the speed of levering up and a parameter $\theta_-$ for the speed of deleveraging. By allowing such asymmetric specification, we find that the dynamics becomes more stable as $\theta_-$ is reduced. The qualitative behavior of the system, namely the existence of stable, locally unstable and globally unstable regimes, is preserved.}

\begin{figure}
        \centering
        \includegraphics[width=0.7\textwidth]{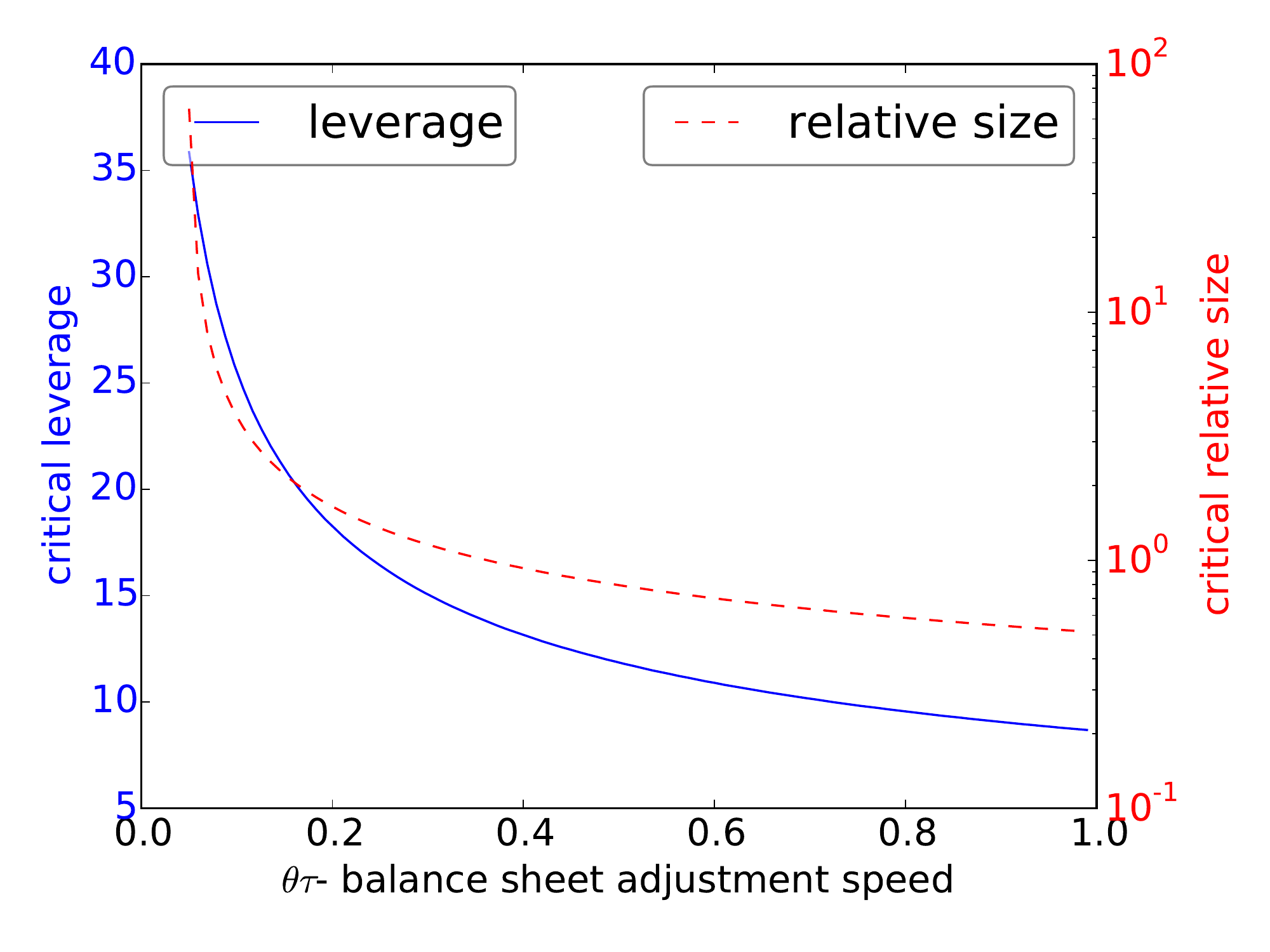}
         \caption{Critical leverage $\lambda^*_c$ (solid blue line, left vertical axis) and the critical value of the relative size of the bank to the fund $R_c(x^*)$ (dashed red line, right vertical axis) as a function of the balance sheet adjustment speed $\theta \tau$.  Other parameters are as in Table 1.  The stability of the financial system can be dramatically improved by lowering the adjustment speed.} 
         \label{FIG::stab}
\end{figure}

To test this we study how the critical leverage $\lambda^*_c$ and critical relative size $R_c(x^*)$ depend on the adjustment speed $\theta \tau$ (we vary $\theta$ and hold $\tau$ constant). The relationship is shown in Figure \ref{FIG::stab}, where the critical leverage is shown on the left vertical axis and the critical relative size on the right vertical axis.  As expected, both the critical leverage (left axis, continuous line) and critical relative size of the bank (right axis, dashed line), decrease dramatically as $\theta \tau$ increases.  This suggests that it is possible to dramatically improve the stability of the financial system if institutions adjust to their leverage targets slowly. 
Similarly, this illustrates the dangers of mark-to-market accounting, which can cause balance-sheet adjustments to be too rapid . 

\section{Leverage control policies}\label{SEC::optimal_policy}
What is the optimal leverage control policy?  The mere fact that the endogenous oscillations of prices and volatility depend on the cyclicality parameter $b$, as shown in Figure~\ref{FIG::stability}, suggests that some policies are better than others.  In this section we introduce a procedure for scoring policies and search for the best policy within the family that we have defined.  We find that the optimal policy depends on parameters of the model, and in particular on the market power of banks in relation to the rest of the financial system.  As the market power of banks increases the optimal policy becomes increasingly countercyclical, and in the limit where the banks play a large role in determining prices it approaches constant leverage.

\subsection{Criterion for optimality}

We define an optimal leverage control policy as one that maximizes leverage for a given level of risk.  Maximizing leverage is desirable because it means that, for a given level of capital, banks are able to lend more money.  We don't model the real economy here, we simply take it as a given that the ability to obtain credit if needed is desirable for the real economy.\footnote{
There may be circumstances where the real economy might overheat as a result of too much credit.  Nonetheless, we assume that, at a given level of risk, all else equal, the option of being able to obtain more credit is desirable for both borrowers and lenders.  When this is not the case they can simply abstain from giving or receiving credit, in which case risk will automatically be lower.}
From a practical point of view it is difficult to control risk while searching the parameter space.   It is much easier to control the average leverage, systematically sweep parameters and measure the resulting risk.

To measure risk we have the luxury of having a simple model, which we can iterate numerically to generate as much data as we need for statistical estimation.  We can then observe the resulting time series of gains and losses for the bank and measure the level of risk associated with this time series.  Because this is an ex post measurement of risk, we call this the {\it observed risk}.

We now compute the trading gains and losses for the bank. The change in the bank's equity due to fluctuations in the price of the risky asset at time $t+1$ is $\Delta E_\text{B}(t) = n(t) \Delta p(t)$, where $\Delta p(t) = p(t+1) - p(t)$. We then define the equity return as
\begin{equation}
\ell(t) = \log\left( \frac{E_\text{B}(t) + \Delta E_\text{B}(t)}{E_\text{B}(t)} \right).
\end{equation}
Note that this captures both the leverage of the bank and the market return of the risky asset since 
\begin{equation*}
\frac{\Delta E_\text{B}(t)}{E_\text{B}(t)} = \frac{n(t) p(t)}{E_\text{B}(t)}\frac{\Delta p(t)}{p(t)} = \lambda(t) w_\text{B} r(t),
\label{tradingLosses}
\end{equation*}
where $r(t)$ is the market return on the risky asset. As expected, leverage amplifies the gains and losses. The total change in equity includes readjustments toward the equity target, but in fact these are small, and the trading losses above are well approximated by the changes in the equity of the bank between time steps.  For consistency with standard risk estimation we use log-returns rather than the simple return $r(t)$.

\begin{figure}[h]
\begin{center}
\includegraphics[width=2.5in]{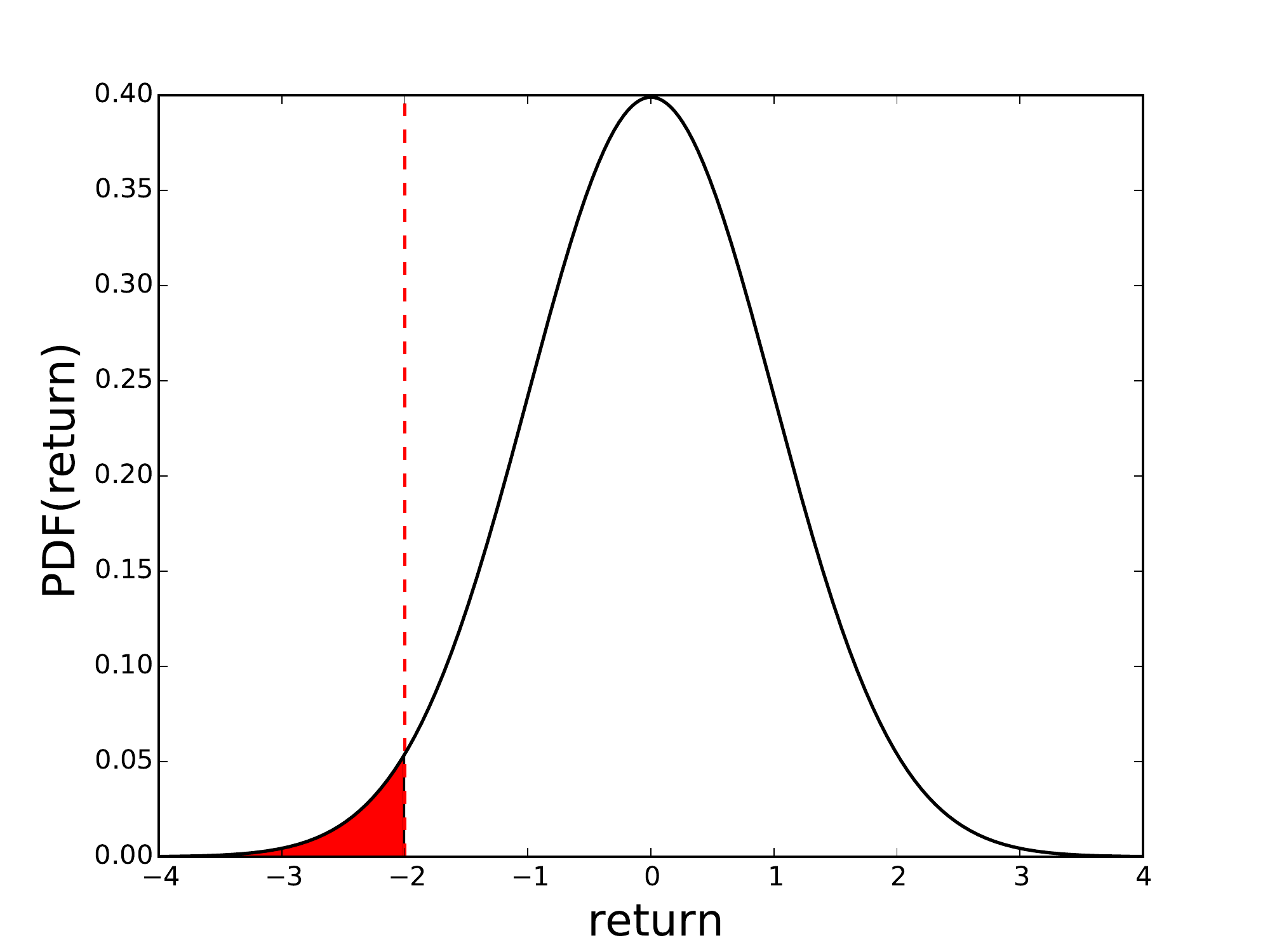}
\end{center}
\caption{\footnotesize{Visual representation of realized shortfall. The solid black lines represents a hypothetical return distribution. The red vertical dashed line is drawn in correspondence of the q-quantile of the distribution. The realized shortfall is the average of the distribution in the red region to the left of the vertical line.}}
\label{RScartoon}
\end{figure}

This then leaves us with the question of how to measure risk.  To do this we follow current thinking as reflected in Basel III and use realized shortfall.  The realized shortfall measures the average tail loss of the bank equity beyond a given quantile $q$.  This is the analog to expected shortfall as used in Basel III, except in this case it is based on the profits and losses realized ex post in the simulation of the model.  It is a measure of the average loss induced by large market crashes, as shown in Figure~\ref{RScartoon}.  The choice of risk metric is not important for the results presented here -- we would get similar results with any other reasonable measure of tail risk. 

For each set of parameters we estimate the realized shortfall using a time average with $T= 5000$ time steps by empirically computing the average loss over the worst $q T$ time observations.\footnote{
For convenience we choose $q$ and $T$ so that their product is an integer.}
Let $\Theta$ be the indicator function with $\Theta(x)=1$ if $x>0$ and zero otherwise, and let $\ell_{q}$ be the threshold loss corresponding to quantile $0 < q < 1$, defined through the relation $\sum_{t=1}^{T} \Theta(\ell_{q} - \ell(t)) =qT$.  The realized shortfall at a confidence level $q$ over a time horizon $T$ is defined as
\be\label{RS}
\text{RS}_q=-\frac{1}{qT}\sum_{t=1}^{T} \ell(t) \Theta(\ell_{q} - \ell(t)).
\ee

\subsection{Balancing microprudential and macroprudential regulation}

To illustrate how the optimal tradeoff between microprudential and macroprudential regulation depends on the properties of the financial system, in this section we investigate three representative scenarios.   The two key properties characterizing the scenarios are the strength of the exogenous clustered volatility and the market power of the banking sector.    The market power of the banking sector is determined by the product of the average relative size $\hat{R}$ of the banking sector, the average leverage $\hat{\lambda}$ and the bank's portfolio weight $w_B$ for the risky asset.   For convenience, to vary the market power of the banking sector we hold  $\hat{\lambda}$ and  $w_B$ constant and vary $\hat{R}$.

\begin{figure}[h]
\centering
\includegraphics[width=0.7\textwidth]{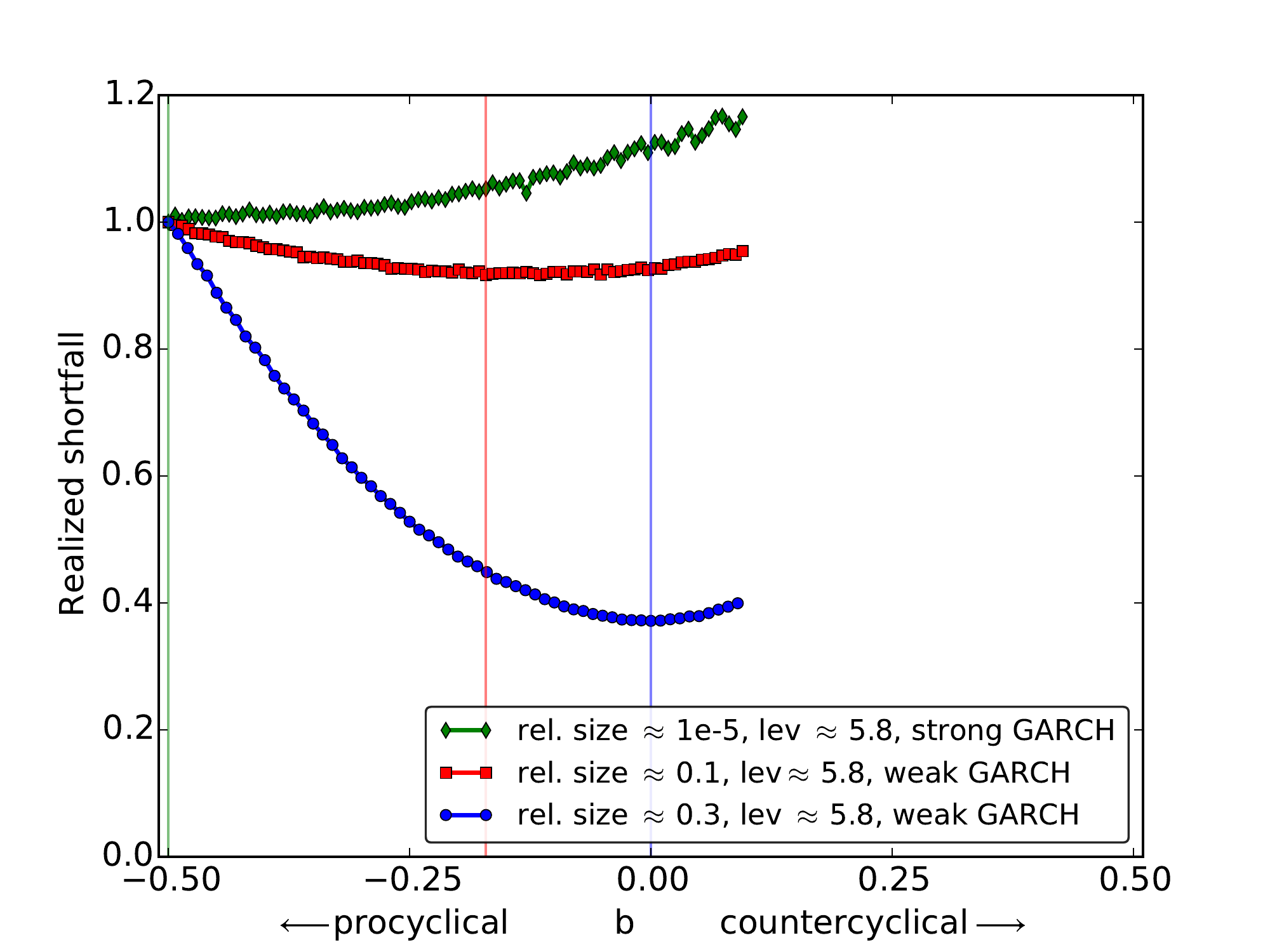} \\
\caption[Determining optimal leverage control policy - calibrated]{\footnotesize{An illustration of how the balance between micro and macroprudential regulation depends on the stability of the financial system.  We plot the observed risk as a function of the cyclicality parameter $b$, investigating three different scenarios corresponding to different levels of clustered volatility and different sizes of the banking sector, with other parameters held constant. Realized shortfall has been normalized by $RS_q(b = -0.5)$ for ease of comparison. (\textbf{Green diamonds})  Microprudential risk dominates when the banking system is relatively small and the exogenous volatility clustering is high.  Not surprisingly, the best policy is $b^* = -0.5$, i.e. Basel II and is strongly procyclical.  (\textbf{Red squares}) There is a mixture of microprudential and macroprudential risk  when the size of the banking sector is increased to $\hat{R} = 0.1$, and the best policy now has $b^* \approx -0.2$, i.e. it is only mildly procyclical.  (\textbf{Blue circles}) Macroprudential risk dominates  when the size of the banking sector is increased still further to $\hat{R} = 0.27$, and the optimal policy has $b^* \approx 0$, i.e. it is very close to constant leverage.  See text for parameters.}}
\label{FIG::b_sweep_lev_const_joint_1}
\end{figure}

In each scenario we sweep the cyclicality parameter $b$ in Equation~(\ref{EQ::leverage_policy}).  This determines the degree of procyclicality or countercyclicality of the leverage control policy.  As we do this we hold the average leverage and the relative size of the banking sector constant at the stated targets, adjusting $\alpha$ and $\overline{E}$ as needed in order to maintain these targets.  We hold all the other parameters of the system constant.\footnote{
Because the instantaneous leverage $\bar{\lambda}(t)$ and the relative size of the banking sector $R(t)$ are emergent properties that vary in time when there is a leverage cycle, controlling them requires some care.  For a given choice of cyclicality parameter $b$ we vary $\alpha$  and $\overline{E}$ to match a target for the average size of the banking sector, $\hat{R} = \frac{1}{T} \sum_{t=0}^T R(t)$, and the average leverage $\hat{\lambda} = \frac{1}{T} \sum_{t=0}^T \bar{\lambda}(t)$.  The leverage is held constant at $\hat{\lambda} = 5.8$ for all scenarios but the size of the banking sector $\hat{R}$ varies as stated.  All other parameters are as in Table \ref{TAB::param_overview} unless otherwise noted.} 
We then measure the observed risk as a function of $b$ and look for a minimum, corresponding to the optimal policy.  The results are shown in Figure~\ref{FIG::b_sweep_lev_const_joint_1}. 

We investigate three scenarios, with the results described below:
\begin{enumerate}
\item {\it Microprudential risk dominates. (Green diamonds)}  This occurs when there is strong exogenous clustered volatility and weak bank market power.  To illustrate this we set the GARCH parameters for strong clustered volatility ($a_0 = 0.001$, $a_1 = 0.04$, $b_1 = 0.95$) and make the banking sector small ($\hat{R} = 10^{-5}$). In this case there is essentially no systemic risk.  The dynamics are dominated by the exogenous volatility, which the historical volatility estimator does a good job of predicting.  Not surprisingly, the best leverage control policy is very close to $b = -0.5$, i.e. it corresponds to Basel II.\footnote{
Note that observations for this scenario only extend up to $b \approx 0.1$ as for larger values of $b$ there exists no model solution with the required output targets for relative size and leverage.}
\item {\it Compromise between microprudential and macroprudential risk. (Red squares) }  This occurs when there is weaker exogenous clustered volatility and intermediate bank market power.  To illustrate this we set the GARCH parameters for weaker clustered volatility ($a_0 = 0.001$, $a_1 = 0.016$, $b_1 = 0.874$) and increase the relative size of the banking sector to $\hat{R} = 0.1$.  The larger size of the banking sector makes the financial system more prone to endogenous oscillations and the risk is minimized for $b^* \approx 0.2$.  This corresponds to a leverage control policy that is still procyclical but is nonetheless closer to constant leverage.  
\item {\it Macroprudential risk dominates (Blue circles).}  This occurs when there is weaker exogenous clustered volatility and strong bank market power.  To illustrate this we set the GARCH parameters as in the previous case but increase the relative size of the banking sector still further to $\hat{R} = 0.27$.  In this case the system becomes very prone to endogenous oscillations and the risk is minimized for $b^* \approx 0$, i.e. using a leverage control policy that is very close to constant leverage.
\end{enumerate}

These three scenarios show that the key determinant of the degree to which micro vs. macroprudential regulation is required is the market power of the banking sector.  As this increases the system becomes more prone to oscillation and therefore more susceptible to systemic risk.  The dynamics emerge because of the tension between the stabilizing influence of the fundamentalist and the destabilizing influence of the banking sector.  As the latter increases in market power a higher degree of macroprudential regulation is required.

The balance between micro and macroprudential risk can be stated in simple terms as a competition between exogenous vs. endogenously generated volatility.   Increasing the size of the banking sector increases the endogenous volatility and means that the system requires a higher level of macroprudential regulation.  This is obvious in the model, but of course in the real world it is hard to tell who is generating volatility and therefore difficult to distinguish the two.  Nonetheless, the market power of the banking sector can potentially be estimated by regulators and provides an important systemic risk indicator.

\section{Conclusion}

\subsection{Summary}
In this paper we have investigated the effect of risk-based leverage policies on financial stability, developing an extension of the dynamical model of leverage cycles introduced by \cite{Aymanns2014}.  The model is a simple agent-based model that includes a leveraged bank, a fundamentalist unleveraged fund and an outside lender, and can be described in terms of a discrete dynamical system defined by six recursion equations. We have roughly calibrated the model to match basic features of the S\&P500 trajectory prior and following the 2007/2008 crash. The model can be simulated in a stochastic or a deterministic regime.

We consider a one parameter family of leverage control policies that varies between procyclical and countercyclical policies. Under a procyclical policy the bank decreases its leverage when perceived risk is high. In contrast, under a countercyclical policy the bank is allowed to {\it increase} its leverage when perceived risk is high.\footnote{
It is important to note that our definition of policy cyclicality does not refer to the level of a macroeconomic indicator such as the credit-to-GDP ratio or the level of asset prices but is defined solely by the bank's response to changes in perceived risk.  Thus the countercyclical policies proposed in this model differ from alternative policies such as the Basel III countercyclical capital buffer, see for example \cite{FPC_2014}.}

We study the stability of the model for different values of bank leverage and the cyclicality of the leverage policy. We find three different regimes: (1) For low leverage the system is stable and settles into a fixed point. (2) As leverage is increased leverage cycles emerge (though for countercyclical policies ($b>0$) this region is small). (3) As leverage is increased further the system becomes globally unstable. The critical leverage at which cycles emerge depends on whether the system is deterministic or stochastic. In the deterministic case the critical leverage is constant for all values of the cyclicality parameter $b$. In the stochastic case the critical leverage increases with $b$.

In our model both procyclical and countercyclical policies are associated with positive feedback loops that can amplify market fluctuations. When leverage is procyclical, banks reduce their balance sheets when perceived volatility is high, pushing prices down and increasing volatility even further, which can trigger a crash. When leverage is countercyclical, in contrast, high volatility leads to higher leverage. This increases the demand for assets, which pushes up both prices and volatility. This positive feedback loop can also be destabilizing.

Our main contribution is the evaluation of different leverage control policies.  Our central result is that the optimal policy depends on the parameters of the financial system.  The key properties are the market power of the bank (which is the product of its leverage, its relative size and its portfolio weight for the risky asset) and the the amplitude of the exogenous clustered volatility.

In the microprudential limit when the market power of the bank is small and exogenous volatility is high, the optimal policy is simply given by Basel II, i.e. Value-at-Risk ($b=-0.5$). As the banking sector becomes larger (either through increasing equity or leverage) the optimal policy becomes less procyclical. In the limit when the bank is very large or highly leveraged, the optimal policy is constant leverage, $b=0$. 

\subsection{Discussion}

Our paper clearly illustrates the interplay between exogenous and endogenous volatility: the microprudential response to exogenous volatility can itself cause endogenous volatility which may dominate over exogenous volatility. This insight is crucial for the effective design of macroprudential policies. Such policies must critically evaluate systemic risks, and make an appropriate tradeoff between micro and macroprudential risk. 

The fact that the degree of cyclicality of the optimal policy depends on the market power of the banking sector suggests that regulation should be flexible to adapt to the financial environment.  The market power depends on both leverage and the size of the banking sector, suggesting that regulation needs to take both of these into account.  Limits on leverage alone are not sufficient.

The most interesting aspect of our model is the sharp and spontaneous emergence of a leverage cycle, even in the deterministic limit.   When we use the industry standard two year moving average to compute historical volatility, and set the other parameters of the model to plausible values, we automatically get a leverage cycle with a period of oscillation in the 10 - 15 year range.  The cycle consists of a slow rise in asset prices and leverage and a slow drop in volatility, which abruptly ends with a crash in asset prices and leverage and a spike in volatility.  The fact that this resembles the Great Moderation and the subsequent crisis is interesting.  It suggests that regulating risk in the style of Basel II might have been the cause of these events, and that the collapse of the housing bubble may have only been the spark that happened to trigger the crash.  Of course we are not claiming that this is true -- there are many other possible causes and the real situation is far too complicated.  We want to stress the word ``suggestive".  We view our model as a cautionary tale, indicating the need to focus more attention on the systemic side effects of risk management.   Our model shows that caution is needed:   The cure can be worse than the disease.

The results of Section~\ref{SEC::adjustment_speed} gives a clear prescription for improved risk management.  We show there that lowering the adjustment speed for leverage targets exerts a strong stabilizing force and can have a dramatic effect on the critical leverage.

This raises the question of whether Basel III has a sufficiently strong limit on leverage and a sufficiently strong capital buffer to prevent a leverage cycle like this from happening in the future.  A research agenda for answering this question might build on our work to make a more realistic model.   Among other things, a more realistic model would allow the bank's portfolio weights to vary, allowing for phenomena like flight to quality; allow the possibly of default; include more realistic aspects of the heterogeneity of the financial system; and implement a more realistic version of the Basel III rules, including risk weights and an asset price-dependent capital buffer.  Such a model could be developed, and we believe it could provide a useful tool for monitoring the stability of the financial system, testing the effectiveness of Basel III, and evaluating possible alternatives.

\section{Acknowledgements}
This work was supported by the European Union Seventh Framework Programme FP7/2007-2013 under Grant
agreement CRISIS-ICT-2011-288501 and by Institute of New Economic Thinking at the Oxford Martin School. CA has also been supported by the German National Merit Foundation and EPSRC (UK). FC acknowledges support of the Economic and Social Research Council (ESRC)
in funding the Systemic Risk Centre (ES/K002309/1).We would like to thank Tobias Adrian and Nina Boyarchenko for useful discussions. The code for the model used in this paper is available upon request.

\bibliography{mybibfile}

\begin{appendix}

\section{Detailed description of the model}\label{APP::Model}

In the following we will describe the modeling details of the assets, the bank, the fund and the market. The model is set in discrete time indexed by $t = \{\tau,2\tau,...,T\tau\}$, where $\tau$ is the length of a time-step.  We leave the length of the time-step as a free parameter in order to allow us to calibrate the model.

\subsection{Assets}\label{SEC::Assets}
Let $p(t)$ be the price of the risky asset at time $t$. The risky asset can be thought of as any traded security such as stocks, asset backed securities or bonds.  We assume that there is exactly one unit of the risky asset which is infinitely divisible. 
The return on the price of the risky asset is $r(t) = \log[p(t)/p(t-\tau)]$.  We denote the fraction of the risky asset held by the bank by $n(t) \in [0,1]$. Since only the bank and the fund can invest in the risky asset, the fraction of the risky asset held by the fund is simply $1-n(t)$.  The risk free asset is analogous to cash.  The price of the risk free asset is constant and equal to one.  

\subsection{Agents}\label{SEC::Agents}
\textcolor{black}{There are two representative agents. The first is a bank, denoted B, and the second is a fund, denoted F.}

\subsection{Bank}\label{SEC::Bank}
\paragraph{Balance sheet}  Assume the bank divides its assets $A_{\text{B}}(t)$ in a fixed ratio $w_{\text{B}}$ between the risky asset and cash $c_B(t)$, so that the banks owns $n(t)$ shares of the risky asset with price $p(t)$.   The relevant accounting relations are:
\begin{equation*}
\begin{aligned}
\text{Risky investment} &= n(t)p(t) = w_{\text{B}} A_{\text{B}}(t),\\
\text{Risk free investment} &= c_{\text{B}}(t) = (1-w_{\text{B}}) A_{\text{B}}(t), \\
\text{Total assets} &= A_{\text{B}}(t) = c_{\text{B}}(t) + n(t)p(t).
\end{aligned}
\end{equation*}
The bank's liabilities $L_{\text{B}}$ have a maturity of one time step and are freely rolled over or expanded. There is no limit to the reduction in $L_{\text{B}}$; in principle the bank could pay back its entire liabilities in one time step.

\textcolor{black}{The bank adjusts its equity toward a fixed target $\overline{E}$.  This guarantees that neither the bank nor the fund asymptotically accumulates all the wealth and makes the long-term dynamics stationary, with only a small effect on the short term dynamics.  The dividends paid out by the bank are invested in the fund and new capital invested in the bank comes from the fund.   If the bank deviates from its equity target $\overline{E}$ it either pays out dividends or attracts new capital from outside investors at a rate $\eta$ to adjust its equity closer to the target, so that its equity changes by}
\begin{equation}
\kappa _\text{B}(t) = \eta (\overline{E}-E_\text{B}(t)).
\end{equation}
Taking both the changes in price and the active adjustments in equity into account, the bank's equity at time $t+\tau$ is
\begin{equation}
E_{\text{B}}(t+\tau) = n(t)p(t+\tau) + c_{\text{B}}(t) - L_{\text{B}}(t) + \kappa _\text{B}(t),
\end{equation}
and the bank's leverage is
\begin{equation}\label{EQ::leverage}
\lambda(t + \tau) = \frac{\text{Total Assets}}{\text{Equity}} = \frac{n(t)p(t+\tau)/w_\text{B}}{E_{\text{B}}(t+\tau)}.
\end{equation}
We assume the bank enforces risk control through a target leverage $\bar{\lambda}(t)$, corresponding to a target portfolio value $\bar{A}_{\text{B}}(t)(t) = \bar{\lambda}(t) E_{\text{B}}(t)$.

\paragraph{Estimation of perceived risk} The bank relies on historical data to estimate the perceived variance of the risky asset $\sigma^2(t)$. To do so the bank computes an exponential moving average of squared returns of the risky asset.  This approach is similar to the RiskMetrics approach, see \cite{Morgan1996}. In particular
\begin{eqnarray}\label{EQ::variance}
\sigma^2(t+\tau) &=& (1 -\tau\delta) \sigma^2(t) + \tau\delta r^2(t) \nonumber \\ 
&=& (1 -\tau\delta) \sigma^2(t) + \tau \delta\left( \log\left[\frac{p(t)}{p(t-\tau)}\right]\frac{t_{\rm VaR}}{\tau}\right)^2,
\end{eqnarray}
where the term $t_{\rm Var}/\tau$ rescales the return over one time-step $\tau$ to the return over the horizon $t_{\rm VaR}$ used in the computation of the capital requirement.
The parameter $\tau \delta \in (0,1)$ implicitly defines the length of the time window over which the historical estimation is performed. We define the typical time $t_{\delta}$ as the time at which an observation made at $t-t_\delta$ has decayed to $1/e$ of its original contribution to the exponential moving average. Thus $t_\delta = -\tau/\log[1-\tau\delta] \approx 1/{\delta}$ for $\tau\delta \ll 1$.

\subsection{Fund investor}\label{SEC::fund}
The fund investor represents the rest of the financial system, and plays the role of a fundamentalist noise trader.   Since the fund is not leveraged its assets $A_{\text{F}}(t)$ are equal to its equity, i.e. $E_{\text{F}}(t) = A_{\text{F}}(t)$.   Just as for the bank, the fund invests $w_{\text{F}}(t)$ of its assets in the risky asset and $1-w_{\text{F}}(t)$ in cash; a key difference is that the fund adjusts its portfolio weight $w_{\text{F}}(t)$ whereas the bank's weight is fixed.  The relevant accounting relations are
\begin{equation*}
\begin{aligned}
\text{Risky investment} &= (1-n(t))p(t) = w_{\text{F}}(t) A_{\text{F}}(t),\\
\text{Risk free investment} &= c_{\text{F}}(t) = (1-w_{\text{F}}(t)) A_{\text{F}}(t), \\
\text{Total assets} &= A_{\text{F}}(t) = c_{\text{F}}(t) + (1-n(t))p(t),
\end{aligned}
\end{equation*}
and the fund's equity is
\begin{equation}\label{EQ::equity_fund}
E_{\text{F}}(t+\tau) = (1-n(t))p(t+\tau) + c_{\text{F}}(t) + \kappa_{\text{F}} (t).
\end{equation}
\textcolor{black}{The fund's cash flow $\kappa _{\text{F}} := -\kappa _{\text{B}} $} mirrors the dividend payments or capital injections of the bank. 

We have already explained the motivation for the fund's demand function in the main text. Here, we simply note that we rescale the deviation of the price of the risky asset to the fundamental by the current price of the risky asset in order to make portfolio weight adjustments independent of the scale of the price of the risky asset. Otherwise, the portfolio weight would likely exceed its natural bounds, i.e. $w_F \in [0,1]$.
In order to introduce heteroskedasticity we make $s^2$ time varying according to a simple GARCH(1,1) process of the form
\begin{equation}
\begin{aligned}
s^2(t) &= a_0 + a_1 \chi^2(t-1) + b_1 s^2(t-1), \\
\chi(t) &= s(t) \xi(t).
\end{aligned}
\end{equation}
When the parameters $a_1$ and $b_1$ are zero the returns $r(t)$ of the risky asset are normally distributed as $n(t) \rightarrow 0$, i.e. as the fund dominates the market, and the price process will resemble a mean reverting random walk with constant volatility.

\subsection{Market mechanism}
The price of the risky asset is determined by market clearing. For this we construct the demand functions for the bank and fund  ($D_{\text{B}}$ and $D_{\text{F}}$ respectively) as follows:
\begin{equation*}
\begin{aligned}
D_{\text{B}}(t+\tau) &= \frac{1}{p(t+\tau)} w_{\text{B}} A_{\text{B}}(t+\tau) \\
		 &= \frac{1}{p(t+\tau)} w_{\text{B}} ( n(t)p(t+\tau) + c_{\text{B}}(t) + \Delta B(t)), \\
D_{\text{F}}(t+\tau) &= \frac{1}{p(t+\tau)} w_{\text{F}}(t+\tau) A_{\text{B}}(t+\tau)  \\
		 &= \frac{1}{p(t+\tau)} w_{\text{F}}(t+\tau) ( (1-n(t))p(t+\tau) + c_{\text{F}}(t) ).
\end{aligned}
\end{equation*}
Recall that there is a supply of exactly one unit of the risky asset that is infinitely divisible. We can then compute the market clearing price by equating demand and supply $1 = D_{\text{B}}(t+\tau) + D_{\text{F}}(t+\tau)$. Solving for the market clearing price we obtain
\begin{equation}\label{EQ::clearing_price}
p(t+\tau) = \frac{w_{\text{B}}(c_{\text{B}}(t) + \Delta B(t)) + w_{\text{F}}(t+\tau) c_{\text{F}}(t)}{1 - w_{\text{B}} n(t) - w_{\text{F}}(t+\tau)(1-n(t))}.
\end{equation}
Given the new price we can compute the fraction of the risky asset owned by the bank as follows:
\begin{equation}\label{EQ::new_fraction}
n(t+\tau) = \frac{1}{p(t+\tau)}  w_{\text{B}} ( n(t)p(t+\tau) + c_{\text{B}}(t) + \Delta B(t) ).
\end{equation}

\subsection{Finding the fixed point}
We begin by considering the conditions for a fixed point of the $g(\cdot)$ as defined in Equation \ref{EQ::6Dmap}.\footnote{For the deterministic system it is simple to derive a set of differential equations for the continuous-time limit. We have checked that the qualitative behavior of the system in continuous time is the same as that of the discrete system in this case. For simplicity, and for consistency with Section \ref{SEC::optimal_policy} where numerical simulations for the discrete stochastic case are considered, we present here results for the discrete dynamical system.}
\begin{enumerate}
\item The price is at the noise trader's fundamental value: \\ $p^* = \mu  \ \implies \ w_{\text{F}}(t+\tau) = w_{\text{F}}(t)$.
\item The bank's perceived risk is 0: \\ $\sigma^{2*} = 0 \ \vee \ p(t) = p(t-\tau) = \mu \ \implies \ \sigma^2(t) = \sigma^2(t+\tau)$. 
\item The bank is at its target leverage consistent with $\sigma^{2*} = 0$: \\ $ \lambda^* = \frac{A^*_\text{B}}{A^*_\text{B}-L^*_\text{B}} = \bar{\lambda}(t) = \alpha(\sigma^2_0)^b \ \implies \ \Delta B(t) = 0$.
\item The bank is at its target equity: \\ $E^*_\text{B} = A^*_\text{B} - L^*_\text{B} = \overline{E} =  \ \implies \ \Delta E_{\text{B}}(t) = 0$.
\item The bank's ownership of the risky asset is consistent with the price, leverage target and equity target at the fixed point: \\ $n^* = \lambda^* E^*_\text{B} w_{\text{B}} / \mu$.
\end{enumerate}
The fixed point is therefore:
\begin{equation}\label{EQ::fixed_point}
\begin{aligned}
x^* &= (\sigma^{2*},w_{\text{F}}^*,p^*,n^*,L^*_\text{B},p'^*) \\
	&= (0,w_{\text{F}}(0),\mu, \frac{1}{\mu} \alpha\sigma^{2b}_0\overline{E}w_{\text{B}}, (\alpha\sigma^{2b}_0 - 1)\overline{E},\mu),
\end{aligned}
\end{equation}
where we picked $w_{\text{F}}^* = w_{\text{F}}(t=0)$, the initial value of the fund's investment weight, since at $p^* = \mu$ any $w_{\text{F}}$ will remain unchanged. Since $w_{\text{F}}^*$ is not specified by the fixed point condition, there is essentially a set of fixed points for $w_{\text{F}} \in [0,1]$. As such it is useful to interpret $w_{\text{F}}^*$ as a parameter of the model determined by an appropriate initial condition. We choose $w_{\text{F}}(0) = 0.5$ throughout.

We can distinguish two cases for the existence of the fixed point. In the first case $\sigma^2_0 = 0$. In this case, provided that $b<0$, the fixed point has an infinity in $n^*$. This fixed point has no economic meaning. Therefore, we will restrict our analysis to the case where $\sigma^2_0 > 0$ in which the fixed point is well defined.

In the case studies in Figure \ref{FIG::example1_2} and Figure \ref{FIG::example3_4} in Section \ref{SEC::dynamcis} we saw that the properties of the system dynamics depended heavily on the relative proportions of the fund versus the bank as this determines the impact of the bank on the price of the risky asset. Therefore, before moving on we define the relative size of the bank to the fund at the fixed point as:
\begin{equation}\label{EQ::rel_size}
\begin{aligned}
R(x^*) &= \frac{A^*_\text{B}}{A^*_\text{F}} 
	   = \frac{\lambda^* E^*_\text{B}}{(1 - n^*)p^*/w_{\text{F}}^*}  &=  \left( \frac{\mu}{\overline{E}} \frac{1}{\alpha \sigma_0^{2b} w_{\text{F}}(0)} - \frac{w_{\text{B}}}{w_{\text{F}}(0)} \right)^{-1}
\end{aligned}
\end{equation}
Clearly, as the equity of the bank goes up, its size relative to the noise trader will increase. Similarly if the bank risk parameter $\alpha$ or the risk off set $\sigma^2_0$ is increased, the bank's leverage at the fixed point will increase whereby its size relative to the fund will increase. 

\subsection{Existence of critical leverage and bank riskiness}
In order to assess the stability of the fixed point we compute the Jacobian matrix $J_{ij} = \partial g_i / \partial x_j$. We then evaluate the Jacobian at the fixed point $x^*$ and compute the eigenvalues $e_i$ of the corresponding matrix. In this particular case the eigenvalues cannot be found analytically. Instead, we compute the eigenvalues numerically using the parameters specified in Table \ref{TAB::param_overview}. With the help of the eigenvalues we can distinguish between local stability and instability of the fixed point. If the absolute value of the largest eigenvalue $|e_{+}| > 1$ the system exhibits chaotic oscillations, while it is locally stable if $|e_{+}| < 1$. We assess the global stability of the system via numerical iteration of the map in Equation \ref{EQ::6Dmap}.

\begin{figure}
\centering
\includegraphics[width=0.7\textwidth]{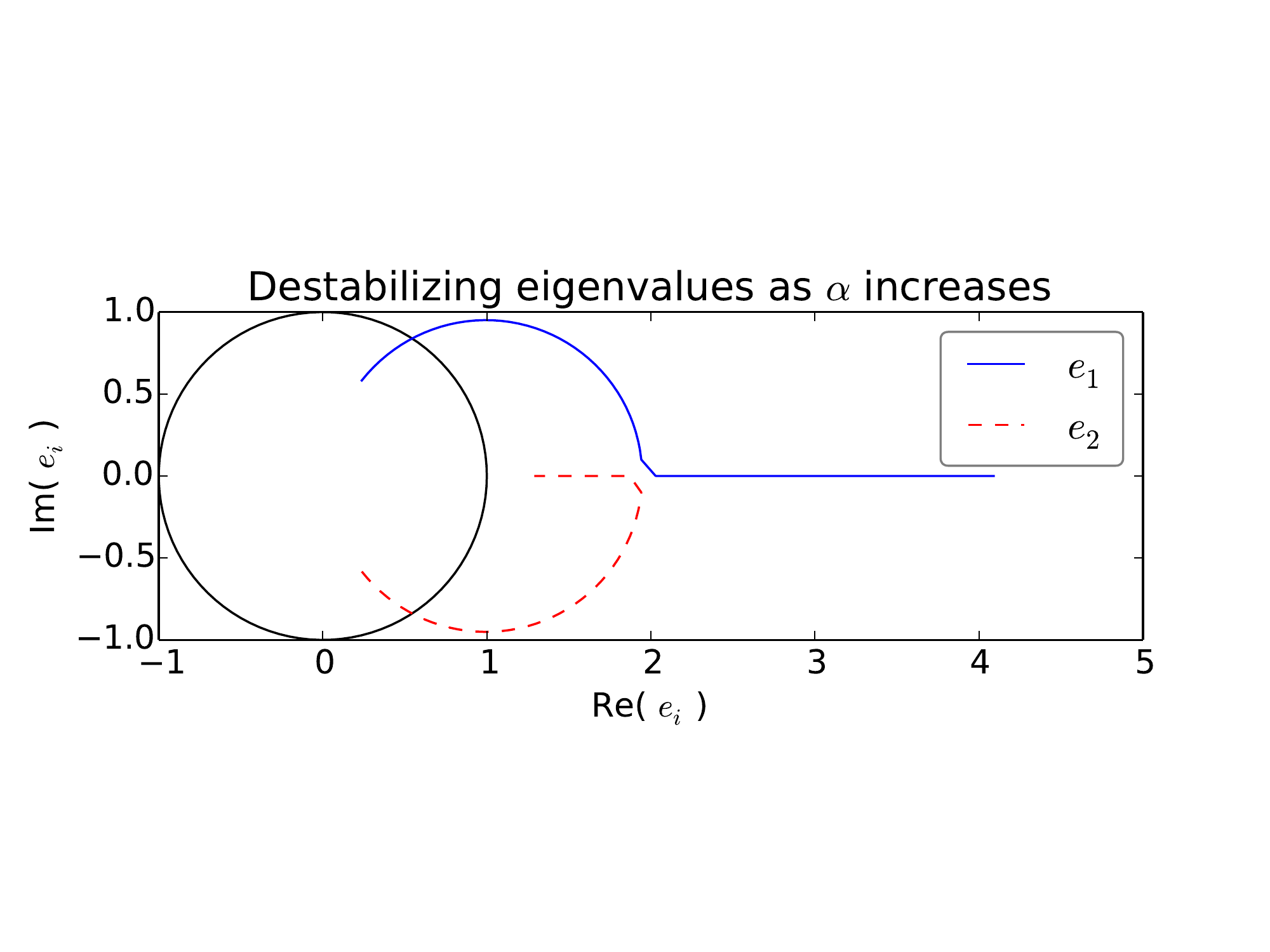}
 \caption{Numerical evaluation of the destabilizing eigenvalues (specific results will depend on parameter choice) for different values of $\alpha$. The destabilizing eigenvalues are the two largest eigenvalues that first cross the unit circle from within.} 
         \label{FIG::eig}
\end{figure}

Now, suppose we increase the bank risk parameter $\alpha$ and study how the eigenvalues of the Jacobian change while keeping all other model parameters constant. It is clear from Equation \ref{EQ::rel_size} and Equation \ref{EQ::fixed_point}, that as we increase $\alpha$ we will increase both the bank's leverage and relative size to the fund at the fixed point. We therefore expect that firstly the bank's market impact increases and secondly that the bank becomes more fragile due to its increased leverage. Thus overall, we expect that if we increase $\alpha$ sufficiently, we should observe a transition from the fixed point dynamics to leverage cycles. We summarize the evolution of the two largest eigenvalues of the Jacobian in the complex plane in Figure \ref{FIG::eig}. The eigenvalues start out at a point within the unit circle on the complex plane (i.e. $|e_{i}| < 1$). Then as $\alpha$ is increased the magnitude of the eigenvalues increases. The critical bank riskiness $\alpha_c$ at which the eigenvalues cross the unit circle, corresponds to the point at which leverage cycles emerge. Since we keep all other parameters constant, this critical bank riskiness also corresponds to a critical leverage and a critical relative size of the bank to the fund. In particular
\begin{equation}
\begin{aligned}
\lambda^*_c &= \alpha_c \sigma_0^{2b},\\
R_c(x^*) &= \frac{\lambda_c^* E^*_\text{B}}{(1 - n^*)p^*/w_{\text{F}}^*}.
\end{aligned}
\end{equation}

\end{appendix}

\end{document}